\documentclass[sigconf,10pt]{acmart}

\newif\ifACM
\ACMtrue

\usepackage[T1]{fontenc}
\usepackage[utf8]{inputenc}

\usepackage{upquote}

\usepackage{microtype}
\UseMicrotypeSet[protrusion]{basicmath} 

\usepackage{environ}
\usepackage{graphicx,grffile}
\usepackage{multirow}
\usepackage{xspace}
\usepackage{tabularx}
\usepackage{ragged2e}
\usepackage{booktabs}
\usepackage{paralist}
\usepackage[american]{babel}
\usepackage[shortlabels]{enumitem}
\setlist{nosep,leftmargin=*}
\usepackage{courier,color,wrapfig}
\usepackage{balance}
\usepackage{epstopdf}
\usepackage{url}
\usepackage{pifont}
\usepackage{subcaption}

\usepackage{mathtools}
\usepackage[normalem]{ulem}

\usepackage[ruled,vlined,linesnumbered]{algorithm2e}
\usepackage{etoolbox}
\usepackage[textsize=small]{todonotes}

\usepackage[compact,small]{titlesec}

\usepackage{caption}
\usepackage[export]{adjustbox}

\usepackage[capitalize]{cleveref}
\crefformat{section}{§#2#1#3}
\crefformat{appendix}{§#2#1#3}

\usepackage{tikz}
\usetikzlibrary[shapes.misc, shapes.geometric, shapes.arrows, positioning, decorations.pathmorphing, matrix, graphs, fit, calc, arrows, arrows.meta, backgrounds]

\usepackage{tcolorbox}
\usepackage{csvsimple}
\usepackage{adjustbox}

\makeatletter
\def\maxwidth{\ifdim\Gin@nat@width>\linewidth\linewidth\else\Gin@nat@width\fi}
\def\maxheight{\ifdim\Gin@nat@height>\textheight\textheight\else\Gin@nat@height\fi}
\makeatother
\setkeys{Gin}{width=\maxwidth,height=\maxheight,keepaspectratio}
\makeatletter
\g@addto@macro{\UrlBreaks}{\UrlOrds}
\makeatother



\newcommand\paraspace{\vspace*{0.5ex}}
\providecommand\parab[1]{\paraspace\noindent\textbf{#1}}

\apptocmd\normalsize{%
\abovedisplayskip=5pt
\abovedisplayshortskip=5pt
\belowdisplayskip=5pt
\belowdisplayshortskip=5pt
}{}{}

\newcommand{\sysname}{GS-NFS\xspace}
\newcommand{\dgs}{4DGS\xspace}
\newcommand{\gs}{3DGS\xspace}
\newcommand{\vthree}{V\textsuperscript{3}\xspace}
\newcommand{\dpsnr}{$\Delta$PSNR\xspace}
\newcommand{\rcr}{RCR\xspace}

\newcommand{\etc}{\emph{etc.}\xspace}
\newcommand{\ie}{\emph{i.e.,}\xspace}

\newcommand\ckmark[1]{\ding{52}}

\tcbuselibrary{skins, breakable}
\newtcolorbox{outline}[1][]
{
  enhanced,
  breakable,
  skin first=enhanced,
  skin middle=enhanced,
  skin last=enhanced,
  colback=blue!5!white, 
  colframe=blue!50!black, 
  fonttitle=\bfseries, 
  boxrule=0.5pt,
  arc=4mm, 
  left=5pt,
  right=5pt,
  title={Outline},
  #1
}
\tcbset{before upper=\setlength{\parskip}{1ex}}

\begin{document}

\begin{abstract}
  Dynamic 3D Gaussian Splatting (\gs{}) holds great promise as a 3D video streaming technology since it can represent complex 3D scenes with high fidelity. 
  In this approach, every frame in a 3D video represents the environment as a collection of Gaussians with position and other attributes such as scale, rotation, opacity, and color. 
  Frames capture fine details, permit views from any arbitrary perspective, but are an order of magnitude, or more, larger than 2D video frames.
  A line of recent work has explored how to compress dynamic \gs{} frames, but these approaches are often slow, in part because their compression techniques are not amenable to efficient acceleration. 
  \sysname{} accelerates dynamic \gs{} compression and decompression on a GPU, to the point where it can encode and decode at full frame rate. 
  It achieves this by developing novel GPU-based parallelizations of existing algorithms for encoding both positions and attributes of Gaussians.
  As a result, it is 1-2 orders of magnitude faster than the state-of-the-art in encoding and decoding a frame, while offering competitive compression performance and rendering quality.
\end{abstract}

\title{\sysname: Bandwidth-adaptive Streaming of Dynamic Gaussian Splats and Point Clouds}
\titlenote{Patent Pending.}
\author{\texorpdfstring{\fontsize{13pt}{16pt}\selectfont}{}Rajrup Ghosh \quad Haodong Wang \quad Haoran Hong \quad Eduardo Pavez \quad Amartya Chaudhuri \quad Weiwu Pang \quad Harsha V. Madhyastha \quad Antonio Ortega \quad Ramesh Govindan}
\affiliation{
  \institution{University of Southern California}
}
\email{rajrupgh@usc.edu}
\date{}



\ifACM
\renewcommand\footnotetextcopyrightpermission[1]{} 
\setcopyright{none}
\settopmatter{printacmref=false, printccs=false, printfolios=true}
\acmDOI{}
\acmISBN{}
\acmConference[Submitted for review]{}
\acmYear{2018}
\acmPrice{}
\pagestyle{plain}
\fi

\ifACM


\fi

\newcommand{\grantack}[1]{
  \ifthenelse{\equal{#1}{CTA}}{\thanks{Research was sponsored by the Army Research Laboratory and was accomplished under Cooperative Agreement Number W911NF-09-2-0053 (the ARL Network Science CTA). The views and conclusions contained in this document are those of the authors and should not be interpreted as representing the official policies, either expressed or implied, of the Army Research Laboratory or the U.S. Government. The U.S. Government is authorized to reproduce and distribute reprints for Government purposes notwithstanding any copyright notation here on.}
    }{}
  \ifthenelse{\equal{#1}{CRA}}{\thanks{Research reported in this paper was sponsored in part by the Army Research Laboratory under Cooperative Agreement W911NF-17-2-0196. The views and conclusions contained in this document are those of the authors and should not be interpreted as representing the official policies, either expressed or implied, of the Army Research Laboratory or the U.S. Government. The U.S. Government is authorized to reproduce and distribute reprints for Government purposes notwithstanding any copyright notation here on.}}{}
  \ifthenelse{\equal{#1}{Conix}}{\thanks{This work was supported in part by the CONIX Research Center, one of six centers in JUMP, a Semiconductor Research Corporation (SRC) program sponsored by DARPA.}}{}
  \ifthenelse{\equal{#1}{NSFAvail}}{\thanks{This material is based upon work supported by the National Science Foundation under Grant No. 1705086}}{}
  \ifthenelse{\equal{#1}{Conix}}{\thanks{This work was supported in part by the CONIX Research Center, one of six centers in JUMP, a Semiconductor Research Corporation (SRC) program sponsored by DARPA.}}{}
  \ifthenelse{\equal{#1}{CPSSyn}}{\thanks{This material is based upon work supported by the National Science Foundation under Grant No. 1330118 and from a grant from General Motors.}}{}        
  \ifthenelse{\equal{#1}{NeTSLarge}}{\thanks{This material is based upon work supported by the National Science Foundation under Grant No. 1413978}}{}         \ifthenelse{\equal{#1}{NeTSSmall}}{\thanks{This material is based upon work supported by the National Science Foundation under Grant No. 1423505}}{}
}


\maketitle

\section{Introduction}
\label{s:introduction}

Video over the Internet has effected societal and economic transformations.
A technology with potentially similar transformative power, 3D video, is on the horizon.
In 3D video, each frame depicts a representation of a scene in three dimensions, and viewers can view the scene from any perspective.
This technology will enable entirely new ways of viewing sporting events, and new modes of instruction delivery, as well as new visual story-telling paradigms.

\subsection{Gaussian Splatting}\label{sec:3d-gauss-splatt}

%
At the core of any 3D video technology is a technique to represent a 3-D scene.
While many representations exist, 3-D Gaussian Splats~\cite{kerbl3Dgaussians} (or \gs{}) have been the subject of recent research interest.
\gs{} represents a scene as a collection of 3-D Gaussians.
Each Gaussian represents a small, ellipsoidal volume, and is defined by its position in 3-D and a set of attributes.
The Gaussians and their attributes are learned by training on a set of camera images capturing a scene from multiple perspectives.
For this reason, we refer to the set of Gaussians and their attributes as a \gs{} model.

More precisely, a \gs{} model is a collection of $N$ 3D Gaussians, each with the following \textit{attributes}:
\begin{itemize}
\item Mean ($\in \mathbb{R}^3$) represents the position of the Gaussian in 3D space (\textit{i.e.}, the center of the ellipsoid).
\item Scales ($\in \mathbb{R}^3$) define the size of the ellipsoid along its three axes.
\item Rotation ($ \in \mathbb{R}^4$) defines the orientation of the ellipsoid in 3D space, typically represented as a quaternion.
\item Opacity ($ \in \mathbb{R}$) represents the transparency of the Gaussian.
\item Colors: In \gs{}, spherical harmonics (SH), which are functions defined on a surface, represent view-dependent color, capturing realistic lighting effects, material properties, and reflections.
Each SH function is represented by three coefficients corresponding to three colors (RGB).  
The coefficients of a degree $d = 0$ SH---the DC coefficients---represent the overall color of the Gaussian.
Higher degree ($d > 0$) AC coefficients capture more complex lighting and shading effects.
A common choice of SH degree is 3, which results in $48$ SH coefficients per Gaussian.
\end{itemize}
Given a \gs{} model and a viewpoint, a rendering algorithm~\cite{kerbl3Dgaussians} produces an image of the view of the scene from the given perspective.
While training a \gs{} model can take minutes, rendering only requires milliseconds.

\begin{figure*}[t]
  \centering
  \includegraphics{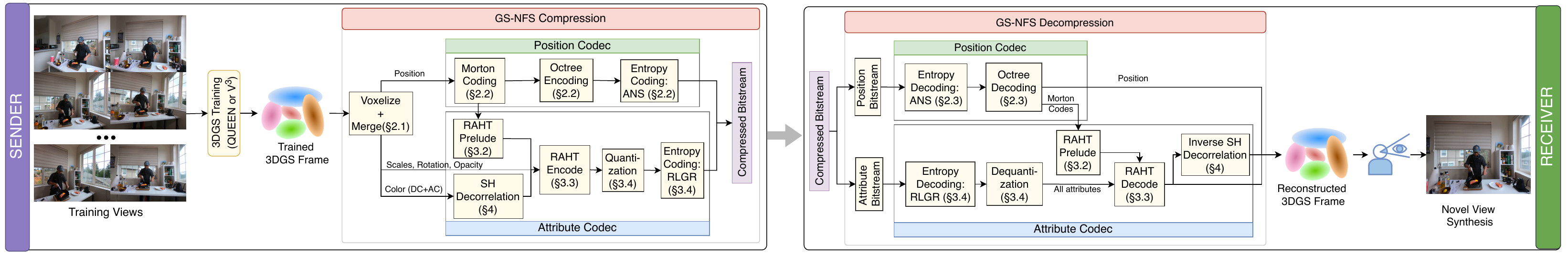}
  \caption{\sysname Architecture. Blocks in yellow are GPU-resident components.}
  \label{fig:architecture}
\end{figure*}

\parab{Dynamic \gs{}.}
A sequence of \gs{} frames constitutes a \textit{dynamic} \gs{} video, often referred to as \dgs{} (time being the fourth dimension).
In this paper, we focus on delivery of \dgs{} content.
This delivery is challenging, because each \gs{} frame can be large.
Each Gaussian requires about 236 bytes to represent its attributes.
\gs{} frames with a million Gaussians are not uncommon, so a single frame can be 236~MB in size.
By contrast, a single frame of 4K UHD video requires 24-30~MB, an order of magnitude less.

\subsection{\dgs{} Compression}\label{sec:dgs-compression}

To reduce the size of \dgs{}, recent work has explored compression techniques.
In addition to reducing size, compression can be used to encode \dgs{} at different bitrate levels for streaming, as for 2D video.
Two \textit{complementary} approaches have emerged (\cref{sec:related-work} discusses related work in detail).

\textit{During-training} compression techniques reduce information in a variety of ways.
%
QUEEN~\cite{queen}, 4DGC-Pro~\cite{zheng20254dgcproefficienthierarchical4d}, and CompGS++~\cite{compgsplus} reduce the number of Gaussians by using temporal prediction between key frames and their immediately following frames.
%
A frame following a key frame is represented using \textit{residuals}---differences with respect to the key frame.
Other approaches exploit human tolerance for small color distortions.
Vega~\cite{Vega} compresses SH (color) coefficients aggressively by first grouping semantically-related Gaussians, then training an MLP for each Gaussian group.
QUEEN~\cite{queen} and 4DGC-Pro~\cite{zheng20254dgcproefficienthierarchical4d}  also train a quantizer that can learn optimal quantization levels for SH coefficients. 
Quantization reduces the number of bits required to represent SH coefficients at the expense of information loss.

\textit{Post-training compression} techniques take a sequence of \gs{} frames as input and encode Gaussian attributes using 2-D or 3-D video codecs.
For example, \vthree{}~\cite{v3volumetric} encodes Gaussian attributes in images and compresses the image sequence using a H.264 codec.
MesonGS~\cite{mesongs} and LTS~\cite{LTS} extend point-cloud compression techniques, such as G-PCC~\cite{gpcc} or Draco~\cite{draco}, to compress \dgs{}.
Post-training compression has two advantages: (a) it can be faster than during-training compression (\cite{queen} reports about a 20\% increase in training time with a during-training quantizer), and (b) it can help encode \dgs{} videos at different bitrates without re-training.

\subsection{Approach, Challenges, and Contributions}\label{sec:appr-chall-contr}

\parab{Goal.}
We present \sysname{} (or GS-NFS)\footnote{NFS, or Need for Speed, is a popular car racing game.}, a post-training compression method for \dgs{} which can encode and decode at full frame-rate, 30 frames per second (fps).
It is also mobile-friendly; on a mobile GPU, it can decode some \dgs{} sequences at 25~fps.
No prior work on post-training compression has achieved full frame rate encode and decode.



Full frame-rate decoding is obviously useful; without that, it will not be possible to play \dgs{} videos at frame rate.
We argue that full frame-rate \textit{encoding} is useful for 
three important reasons.
First, for on-demand streaming, it will be necessary to encode \dgs{} videos at different bitrates.
At scale, an efficient encoder can reduce dollar costs significantly for encoding in the cloud.
%
%
Second, today, 2D video providers use per-title \textit{encode optimization}~\cite{netflix_rdo}.
This approach seeks to find the best quality to encode a video, by repeatedly encoding and decoding frames at different qualities.
A fast encoder/decoder like \sysname{}'s can enable similar optimizations for \dgs{}.
Third, recent work in feed-forward \gs{} networks~\cite{zhou2024gpsplus} can generate Gaussian frames within tens of milliseconds.
This can generate Gaussian frames at nearly full frame rate, and \sysname{} can help encode and transmit these frames in real-time.



\parab{Approach.}
To achieve this goal, \sysname{} uses a compression approach similar to a point-cloud compression technique, G-PCC~\cite{gpcc}.
This is a more natural starting point than the 2D compression used in \vthree{}~\cite{v3volumetric}, since a \gs{} frame's representation resembles that of a point-cloud.
The latter contains a set of points, each with one or more attributes; \gs{} is similar, except that each Gaussian has more attributes (\cref{sec:3d-gauss-splatt}).
Next, we describe the G-PCC pipeline for encoding \gs{} frames; MesonGS~\cite{mesongs} has used such an approach.

G-PCC~\cite{gpcc}'s encoder takes a set of positions and associated attributes as input, then serializes (produces a bitstream) this set for transmission.
Its decoder reconstructs the positions and attributes from the bitstream.
To do this, G-PCC (a) \textit{encodes} positions and associated attributes in such a manner that they can be decoded efficiently, and (b) compresses these encodings (either with or without information loss) to ensure bandwidth and storage efficiency.

G-PCC encodes position and attributes using qualitatively different approaches.
It represents positions of Gaussians using a spatial data structure, an \textit{octree}.
Usually, \gs{} Gaussians are spatially sparse, so not all parts of the octree are occupied.
G-PCC serializes the octree such that the bitstream only contains information for occupied octree parts.

For attributes, G-PCC provides several alternatives.
We use Region-Adaptive Hierarchical Transform (RAHT~\cite{raht,de2016compression}), which organizes the Gaussians into a hierarchy of levels and predicts attributes at finer levels from attributes at coarser levels.
The output of RAHT is a set of coefficients, which can be \textit{quantized} (this introduces loss) and then entropy-coded (a lossless step that removes redundancy).

G-PCC's decode pipeline inverts these operations, de-serializing Gaussians from the encoded bitstream.
Unfortunately, G-PCC and MesonGS have high encode and decode times (\cref{sec:baseline-comparisons}).
\sysname{} achieves frame-rate encode/decode by \textbf{\textit{running the entire encoding and decoding pipeline on a GPU}}.
Beyond accelerating encoding, executing the encoder on a GPU is efficient when the encoder takes input from a training method that emits \gs{} frames.
These frames are already in GPU memory, since \gs{} training requires a GPU, thereby avoiding memory copy costs.

\parab{Challenges and Contributions.}
Unfortunately, accelerating G-PCC's algorithms on a GPU is non-trivial.
Unlike 2D video frames, where pixels exhibit regularity, \gs{} training can produce irregularly-spaced Gaussians.
Mapping this to a single-instruction, multi-threaded (SIMT) GPU is difficult.

CPU-resident algorithms for octree encoding and RAHT manage this irregularity by imposing a spatial hierarchy on the data.
Unfortunately, GPUs are a poor fit for computations on hierarchical data.
Traversing the hierarchy requires following pointers; this \textit{pointer chasing} is known to be highly inefficient on a GPU, since if the referenced memory is not in the cache, all warps in a GPU thread must wait for the data to be loaded (sometimes from relatively slow DRAM), resulting in poor performance.
This problem is well-known for GPU-based computations on irregular data in general~\cite{pointer_chasing1,pointer_chasing2}, and for parallelizing point-cloud algorithms in particular~\cite{groot,octree-gpu2011}.

Against this backdrop, \sysname{} is, to our knowledge, the first post-compression encoder/decoder for \dgs{} in which \textit{all} components (\cref{fig:architecture}) execute on a GPU and can encode and decode at full frame rate.
To achieve this, the paper makes the following novel contributions:
\begin{itemize}
\item A GPU-based parallelization of octree encoding (\cref{sec:gpu-octree}).
\item A GPU-based parallelization of RAHT (\cref{sec:gpu-raht}).
\item A technique to improve the compression of \dgs{} (\cref{sec:klt}). \end{itemize}

Our evaluations (\cref{sec:evaluation}) on two different popular \dgs{} datasets demonstrate that \sysname{}: (a) can encode an order of magnitude faster than the state-of-the-art; (b) dominates MesonGS~\cite{mesongs}, a GPCC~\cite{gpcc-adaptive} variant modified for \dgs{}, and LTS~\cite{LTS} both in quality and compression performance, and performs comparably or better than \vthree{}~\cite{v3volumetric} on large scenes; (c) can decode \dgs{} at up to 25~fps on a Jetson Orin; and (d) can encode and decode point clouds and live \dgs{} videos at full frame rate, a capability that, to our knowledge, has not been demonstrated in the literature.

\section{GPU-Accelerated Octree Encoding/Decoding}\label{sec:gpu-octree}

We first describe our GPU-accelerated octree encoding algorithm, which encodes Gaussian position attributes efficiently.

\subsection{Background}\label{sec:background}

\parab{Voxelization.}
Gaussians can be positioned irregularly in space, and all 3-D compression algorithms first \textit{voxelize} Gaussian positions to regularize the data.
Voxelization discretizes the cuboid bounding all the Gaussians into a regular grid of $2^J \times 2^J \times 2^J$ cubes, where $J$ is a voxelization parameter which controls the \textit{resolution} of the \gs{} frame.
Then, it maps each Gaussian's 3-D position to one of the cubes, by quantizing the position coordinates to use $J$ bits.
Multiple Gaussians might fall into a single voxel; \sysname{} replaces those Gaussians with a single Gaussian whose position is at the center of the voxel and other attributes are the average of the original Gaussians.\footnote{This may not be ideal for small $J$, but suffices for the range of $J$ we consider in this paper.}

\parab{The Octree Data Structure.}
\begin{figure}[t]
  \centering
  \includegraphics[width=\columnwidth]{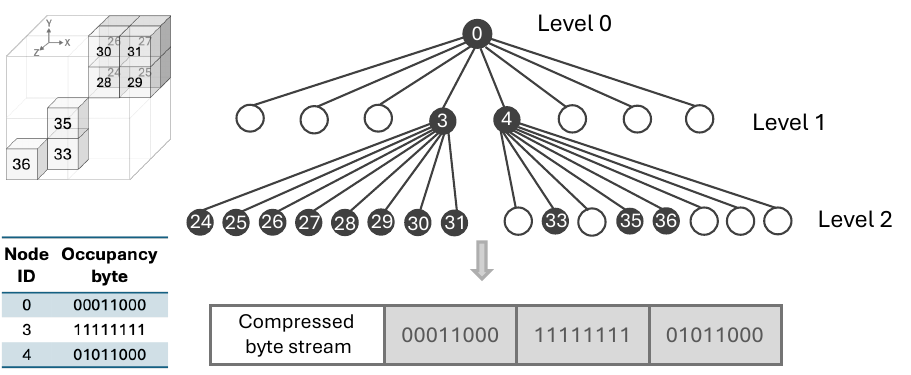}
  \caption{An example of an octree with some occupied voxels.}
\label{fig:octree}
\end{figure}
After voxelization, many voxels are often empty and only a small fraction of voxels are occupied by Gaussians.
An octree~\cite{octree-orig1982} efficiently encodes the positions of occupied voxels to exploit sparsity and spatial locality.
It recursively partitions the 3D space into octants (8 equally-sized child nodes), creating a tree structure where each node represents a cubic region of occupied space.

Consider the octree in~\cref{fig:octree}.
The root node represents the entire 3D space, and its 8 child nodes represent the 8 equally-sized children.
The child nodes that are occupied (\textit{i.e.}, contain at least one Gaussian) can be further subdivided into their own child nodes.
The process continues until we reach a leaf node that contains a single occupied voxel. 

This structure can efficiently encode occupancy.
Each internal tree node uses a byte, where each bit indicates whether the corresponding child node is occupied (1) or empty (0).
For example, if the fourth and fifth bits are set to 1, it means that the fourth and fifth child nodes are occupied, while the rest are empty (\cref{fig:octree}).

\parab{Recursive Octree Encoding.}
3-D compression must \textit{construct} the octree (\textit{i.e.}, determine octants and their occupancy), then \textit{encode} (\textit{i.e.}, serialize) the octree into a bitstream.
Both of these can be accomplished with a single sequential \textit{level-order} traversal from the root to the leaves; a reference implementation of G-PCC~\cite{gpcc-github} uses this approach.
This algorithm starts from the root, determines the occupancy of its child octants, then writes out the occupancy byte of the root.
It then recursively repeats the process for each child in a fixed sequence (\textit{i.e.}, if children are numbered $0..7$, it visits the children in that order), and stops when all occupied voxels form the leaves.

The output of this algorithm is a sequence of bytes representing occupancy at various nodes of the octree.
The total number of bytes in this sequence is proportional to the number of occupied voxels $\hat{N}_v$, which is often more compact than recording the occupancy of all $2^{3J}$ voxels.

In \cref{fig:octree}, for example, only two children of the root node (the 4th and the 5th) contain at least one occupied voxel, so its occupancy byte is $00011000$.
This forms the first byte of the output bitstream.
The second byte corresponds to the root's 4th child, all of whose children are occupied, so its occupancy byte is $11111111$, and so on.

\subsection{GPU-accelerated Octree Encoding}\label{sec:octree-encode-gpu}

\sysname{} parallelizes octree construction and encoding on the GPU.
Before describing this algorithm, we explain why this is a challenge, and how prior work has attempted to accelerate octree construction and encoding.
The next subsection describes GPU-accelerated octree decoding.

\begin{table}[t]
  \centering
  {\scriptsize
  \begin{tabular}{lp{0.7\columnwidth}}
    \hline
    \textbf{Works} & \textbf{Quote}  \\ \hline
    Zhou~\cite{octree-gpu2011} & ``Creating an octree for point clouds directly on the GPU, however, is very difficult, mainly because of memory allocation and pointer creation.'' \\ \hline
    GROOT~\cite{groot} & ``While the tree structure efficiently handles the sparsity of 3D data and subdivides only the volumes where the point exists, the irregularity requires traversing the serialized byte stream and recursively calculating the child node geometry. The exact positions that represent intermediate nodes depend on the occupancy of their ancestors, which cannot be parallelized.'' \\ \hline
  \end{tabular}}
  \caption{Parallelization challenges for octree encoding reported in prior work.}
  \label{tab:octree-challenge}
\end{table}

\parab{Challenges.}
Recursive octree construction is inherently sequential.
State-of-the-art point cloud compression algorithms~\cite{gpcc,draco,pcl} implement sequential CPU-based octree construction and encoding.
Constructing the octree requires traversing the tree top-down and creating nodes and pointers.
This is difficult to efficiently parallelize on a GPU since the memory layout of the tree can be irregular, and tree-traversal might require pointer-chasing, resulting in GPU thread stalls while waiting for memory accesses to complete~\cite{octree-gpu2011,octree-cpu2020,groot}.
Generating the occupancy bitstream also requires a tree traversal in level-order and pointer-chasing to determine the occupancy byte of a node depending on existing children, which can be inefficient on a GPU.
\cref{tab:octree-challenge} includes quotes from prior work~\cite{octree-gpu2011,groot} that illustrate the difficult of parallelizing octree encoding.

\parab{Prior Work on Parallelization.}
One line of work has explored CPU parallelism for octree construction and encoding.
ViVo~\cite{vivo}, MeshReduce~\cite{meshreduce}, and MetaStream~\cite{metastream} use coarse-grained CPU parallelism by partitioning the 3D points into separate blocks, generating the octree encoding independently for each block and concatenating these encodings.
Others~\cite{octree-cpu2020,groot} employ slightly more sophisticated hybrid CPU-GPU strategies.
For example, Groot~\cite{groot} sequentially constructs the octree top-down up to a certain depth $d$, then generates encodings in parallel for each occupied voxel (on a GPU) by traversing the tree bottom-up up to level $d$.

Another line of work~\cite{octree-cpu2008,octree-gpu2011,octree-gpu2012} has achieved GPU-accelerated parallelization of octree construction.
They linearize the positions in 3D to a 1D list using a space-filling curve (the Morton code, described below) while preserving spatial locality~\cite{morton-code-spc}.
Thus, the generated octree is not explicitly represented as a tree structure, but rather as arrays of indices in the space-filling curve for each node at a certain level. 
In these approaches, however, \textit{encoding is still sequential}, since that requires an in-order tree traversal.
\sysname{} uses a similar space-filling curve, but parallelizes both construction and encoding on a GPU.

%


\begin{figure}[t]
  \centering
  \includegraphics[width=0.6\columnwidth]{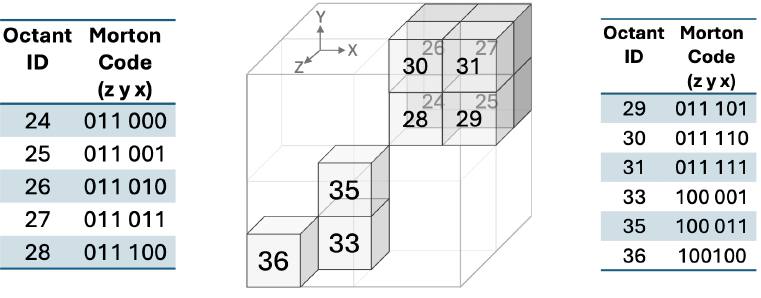}
  \caption{This figure shows the Morton codes of the occupied voxels in \cref{fig:octree}.}
\label{fig:morton}
\end{figure}

\parab{Morton Codes.}
The key to our GPU-accelerated octree encoder is to never build explicit tree pointers.
Rather, we represent each occupied voxel by its \emph{Morton code}~\cite{morton-code-spc}, which is a bit-interleaving of its integer voxel coordinates.
For a voxelized coordinate $\hat{v}=(x,y,z)\in\{0,\ldots,2^J-1\}^3$, the Morton code is a bit string $z_{J-1}y_{J-1}x_{J-1}\ldots z_0y_0x_0$ where $x_b,y_b,z_b$ are the $b$-th least-significant bits of $x,y,z$.

For example, consider the voxelization shown in~\cref{fig:morton}, which depicts a $4\times4\times4$ voxel grid ($J=2$) with 11 occupied voxels.
The occupied voxel with ID $33$ has coordinates $(1,0,2)$ so its Morton code is $100001$.
The Morton code is obtained by first taking the most significant bits of the 3 coordinates ($100$) in the order from $z-x$, followed by the next most significant bits (0 for the z-coordinate value of 2, 0 for the y-coordinate value of 0, and 1 for the x-coordinate value of 1), and so on.

The Morton code has three properties crucial for GPU-acceleration of octree construction, encoding, and decoding.
First, it preserves spatial locality: when we sort the voxels by their Morton code, voxels near each other in space will be near each other in the sorted order~\cite{morton-code-spc}.
For example, if in~\cref{fig:morton}, the voxel with ID $33$ and coordinate $(1,1,2)$ is also occupied, it would have the Morton code $100011$.
This differs from the Morton code of its neighboring voxel $(1,0,2)$ only in the last 3 digits.
This property ensures memory locality if voxels are laid out in GPU memory in Morton order, and avoids non-local memory access overheads.

Second, the Morton code for the parent of a voxel is a prefix of the voxel's Morton code.
In \cref{fig:morton}, the parent of $(1,0,2)$ (ID $33$) has the code $100$, which can be obtained by shifting the voxel's Morton code three bits to the right.
Conversely, to obtain the Morton code of, say, the 5th child of an octree node, we simply append $101$ to the node's Morton code.

Third, given the Morton code for a voxel, we can obtain its voxelized coordinates using the definition of the Morton code.
In our example, given the code $100011$, the first coordinate is obtained by concatenating every third bit starting from position 3, the second coordinate by doing so from position 2, and so on, resulting in the coordinate $(1,1,2)$.


Thus, if given the Morton code for an occupied voxel, it is an $O(1)$ operation to determine its parent.
Conversely, if all octree nodes are sorted in Morton order, then a parent can in an $O(1)$ operation determine its child's Morton code, and look up its occupancy status.
Finally, in $O(1)$, given a node's Morton code, we can obtain its voxel coordinates.

Before explaining how we use these properties to construct, encode and decode octrees, we describe how we parallelize voxelization (\cref{sec:background}).
This is important since voxelization precedes octree construction.
We require all steps in \sysname{} to be GPU-resident to avoid data movement costs (\cref{sec:appr-chall-contr}).

\parab{Voxelization.}
Voxelization (a) discretizes Gaussian positions and (b) merges Gaussian attributes (by averaging them) of Gaussians at the same discretized positions.
The input to this step is an $N \times C$ tensor for a \gs{} frame with $N$ Gaussians, with each Gaussian having $C$ attributes.
Step (a) is simple to parallelize on a GPU by launching one thread per Gaussian, which computes the voxel coordinates (\cref{sec:background}) of the Gaussian.
This step also computes the voxel's Morton code, and creates an array sorted by the Morton code that contains a pointer to the Gaussian.
We use a 64-bit Morton code to support octree depths up to $J=21$, which is sufficient for all scenes of interest~\cite{mesongs}.
If $J\leq 21$, we set to zero the unused most-significant bits in the Morton code.

In this array, Gaussians belonging to the same voxel are contiguous; we identify these \textit{clusters} using standard PyTorch operations.
Then, we launch one GPU thread for each cluster to average the attributes of Gaussians in that cluster.
The output of this process is an $N_v \times C$ tensor, where $N_v$ is the number of occupied voxels.
This is sorted by Morton order, and forms the input to the next step.

\parab{Octree construction.}
In contrast to the recursive top-down approach (\cref{sec:background}),
our construction identifies, \textit{bottom-up}, occupied parents of occupied voxels, and occupied internal nodes up the octree (\cref{fig:octree}).
Let $L_J$ be an array of occupied voxels, sorted by Morton order, obtained directly from the input tensor.
We launch one GPU thread per occupied voxel; this thread computes the parent's Morton code and writes this to a new list $L_{J-1}$.
This array contains parents of occupied voxels in Morton order, but can have duplicates, since multiple occupied voxels can have the same Morton code.
We use CUDA's \texttt{unique} primitive to compact duplicates in parallel, so $L_{J-1}$ has exactly one instance of an occupied parent.

We can now repeat this process to find the next-level occupied parents (\textit{i.e.}, parents of occupied parents) up to the root level.
The output of this process is a list of Morton codes for the occupied nodes at each level, \ie $L_0,\ldots,L_J$, where $L_0$ has the root node, $L_J$ has the leaf nodes, and $L_d$ has the occupied nodes at depth $d$.

\parab{Bitstream generation.}
From these lists, the encoder must emit the octree occupancy bitstream.
Each internal node contributes exactly one occupancy byte, whose $i$-th bit is set if child $i$ exists.
This step requires us to compute the occupancy byte for each internal node.
To compute this, a node must: (a) determine the Morton codes of its children and (b) identify which children are occupied.

Given a parent code $p_m\in L_d$, the Morton prefix of its children can be obtained by $c_m = p_m \ll 3$.
The eight possible children are therefore the contiguous code range $[c_m,c_m+7]$.
To compute the occupancy byte for $p_m$, we need to check which of these eight candidate child codes exist in the child list $L_{d+1}$.
This requires searching for the child codes in $L_{d+1}$.

We make 2 observations that make this search efficient. 
First, all children with Morton codes in the range $[c_m,c_m+7]$ must appear contiguously in the sorted list $L_{d+1}$.
Second, we can perform a binary search for the first occupied child code $\geq c_m$ in $L_{d+1}$, since that listed is sorted.

To compute the occupancy bytes for all octree nodes, we use a \textit{top-down} approach.
We compute occupancy bytes for nodes in $L_d$ before doing so for nodes in $L_{d+1}$.
At level $L_d$, our octree encoder launches a custom CUDA kernel\footnote{For efficiency, when necessary, we replace PyTorch operations with custom CUDA kernels to improve encoding speed.}
with one thread per parent node $p_m$ in $L_d$.
It initializes $p_m$'s occupancy byte to 0.
Then, the thread for $p_m$ performs the following steps:.
(i) compute $c_m = p_m \ll 3$;
(ii) binary-search $L_{d+1}$ to find the first occupied child code $\geq c_m$;
(iii) scan forward over at most 8 entries until the child code exceeds $c_m+7$;
(iv) set bit for every child code $c_m$ found.
The thread then writes its occupancy byte to a known location in the GPU buffer.
We present the detailed algorithm in \cref{algo:octree-encode-gpu}.


At the end, all the occupancy bytes are written in order, similar to \cref{fig:octree}.
To improve compression efficiency, \sysname{} then runs a GPU-based entropy encoder (\cref{fig:architecture}), ANS~\cite{nvidia_nvcomp}.


\subsection{GPU-Accelerated Octree Decoding}\label{sec:octree-decode-gpu}

\parab{Inputs and Output.}
The input to this algorithm is a sequence of occupancy bytes in the octree.
We obtain this from the output of the ANS entropy decoder.
The output of the algorithm is a list of voxelized positions of occupied voxels.
Each such position corresponds to a Gaussian, whose attributes we obtain as described in \cref{sec:gpu-raht}.

This algorithm also needs to know $J$, the depth at which the \gs{} frame was encoded, and the number of occupied nodes $n_d$ at each level $d$ in the tree.
Such information is encoded in the bitstream as \textit{metadata}. 


\parab{Prior work.}
GROOT and~\cite{octree-cpu2020} try to decode lower levels of the octree in parallel on a CPU, but most tree levels are still decoded sequentially on the CPU top-down.
We know of no other work that has GPU-accelerated decoding.

\parab{Octree decoding.}
Our octree decoding algorithm relies on properties of the Morton code.
We note that the first byte is the occupancy byte of the root.
From this, we can determine (a) which children are occupied, and (b) the Morton code of the children (both $O(1)$ operations).
We can repeat this process to obtain the Morton codes of all nodes in the tree, and eventually, the Morton codes of all occupied voxels.
From these, we can obtain the voxelized positions of the occupied voxels using the definition of the Morton code.

To parallelize this, we use a top-down level-synchronous approach (similar to the encoder), where, at each level $d$, we launch one thread per octree internal node.
For level $d$, we can determine from the metadata where the occupancy bytes for that level start and end.
But each node at that level can have a variable number of occupied children.
To determine the offset of the $k$-th occupied node, we first create an array per internal node that counts the number of occupied children (using the population count CUDA primitive), then do a CUDA prefix sum on that array to find the offset of this node's children in the occupancy bytes at level $d+1$.

The $k$-th occupied node writes the Morton codes of its occupied children into a separate array starting at the corresponding offset.
Then, the decoder launches one thread for each occupied thread at level $d+1$, and the process repeats.
After processing all levels, the decoder's output is a list of Morton codes for the leaf nodes, which correspond to the occupied voxels.
The decoder launches a final CUDA kernel to decode the Morton codes back into integer voxel coordinates by reversing the bit interleaving.
\cref{app:sysn-octr-decod} contains a detailed listing of this algorithm.

\section{GPU-Accelerated Attribute Encoding/Decoding}\label{sec:gpu-raht}

In this section, we describe how we GPU-accelerate the encoding and decoding of Gaussian attributes, such as opacity, scale, rotation, and spherical harmonics.

\subsection{Background: RAHT}\label{sec:background-1}

Region-Adaptive Hierarchical Transform (RAHT) is an algorithm for encoding and decoding the attributes of sparse, spatially non-uniform data (such as point clouds~\cite{de2016compression} and \gs{} frames).
The algorithm is complex to explain in its entirety, so we use an example to explain the algorithm; details can be found in~\cite{de2016compression,raht}.
%
To further simplify the description, we describe how it encodes a \textit{single attribute}, say opacity.

\begin{figure}[t]
  \centering
    \includegraphics[width=\columnwidth]{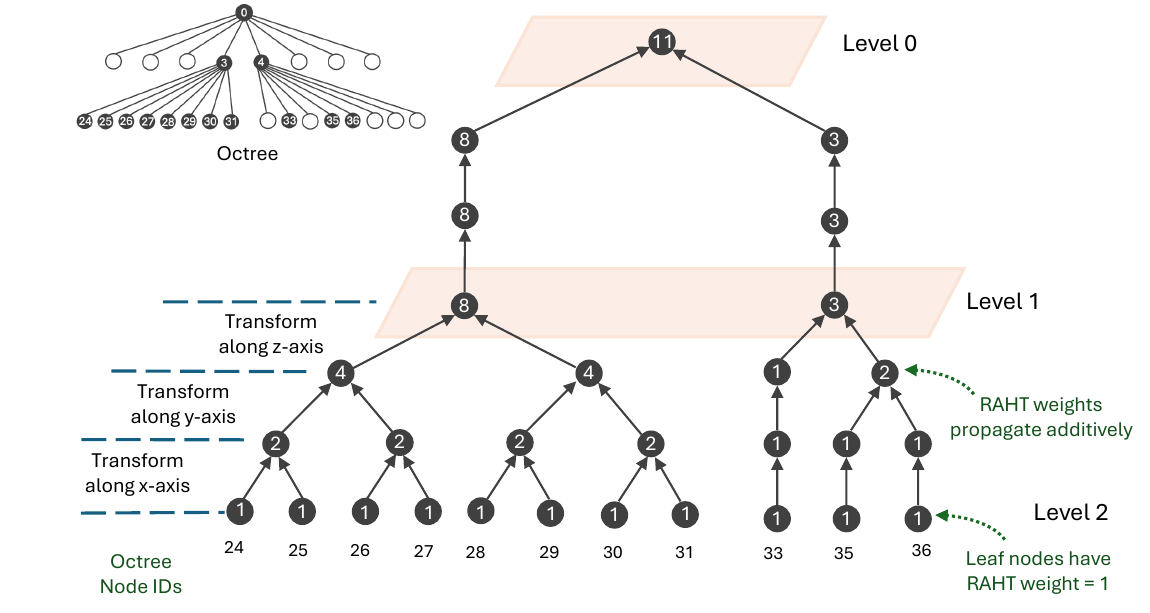}
    \caption{Example to illustrate the RAHT algorithm.}
    \label{fig:raht} 
\end{figure}

RAHT exploits the octree structure derived from voxel positions (\cref{fig:raht}).
It operates on a binary tree derived from the octree; in this binary tree, each octree level is represented by 3 levels in the binary tree, each of which corresponds to one of the directional axes.
The leaves of the tree correspond to voxels, and in \cref{fig:raht} we only show branches in the binary tree leading to occupied voxels, for simplicity.


If a node has a sibling (\textit{e.g.}, nodes \textbf{35} and \textbf{36} in \cref{fig:raht}), RAHT decomposes the attribute into:
\begin{enumerate}
    \item \textbf{Low-frequency Coefficient:} A weighted average representing coarse signal information. 
    \item \textbf{High-frequency Coefficient:} A weighted difference representing the signal details.
\end{enumerate}
More precisely, if two siblings $l$ and $r$ have attribute values $a_l$ and $a_r$, then computing these coefficients requires a $2 \times 2$ matrix-vector multiplication:
\begin{equation}
  \begin{bmatrix}
  a^0\\
  a^1
  \end{bmatrix}
  =
  \frac{1}{\sqrt{w_l+w_r}}
  \begin{bmatrix}
  \sqrt{w_l} & \sqrt{w_r}\\
  -\sqrt{w_r} & \sqrt{w_l}
  \end{bmatrix}
  \begin{bmatrix}
  a_l\\
  a_r
  \end{bmatrix},
  \label{eq:raht-butterfly}
\end{equation}
In this equation, $a^0$ represents the low-frequency (coarse) coefficient, $a^1$ the high-frequency coefficient, and $w_l$ and $w_r$ represents \textit{weights} associated with $a_l$ and $a_r$ respectively.
The weight at any node is the number of occupied voxels in the octant below that node; attributes at leaf nodes have unit weights.
\cref{fig:raht} shows node weights next to the node.
RAHT propagates the low-frequency coefficient to the parent, while storing the high-frequency coefficients at the right sibling.
(If a node does not have a sibling such as node \textbf{33}, it passes its attribute value up to its parent).
RAHT then repeats this procedure level by level, bottom-up, on this tree.

At the end of this computation, the octree root stores the low-frequency component, and internal nodes and leaves store high-frequency coefficients.\footnote{This computation can be intuitively thought of as an extension of the Haar wavelet transform to a sparse 3D grid~\cite{de2016compression}.}
Smoothly varying spatial signals (\textit{e.g.}, point clouds can have smoothly varying color, just like images) have near-zero high-frequency coefficients, especially close to the leaves, which can be quantized and entropy-coded, resulting in significant compression.
In \sysname{}, we use RAHT to encode all \gs{} attributes (\cref{sec:3d-gauss-splatt}) other than positions.
RAHT decoding inverts this operation; for brevity, we discuss only the RAHT encoder and leave the decoder (inverse-RAHT) to~\cref{app:sysn-raht-encod}.

\subsection{Background: A Re-formulated RAHT Algorithm}\label{sec:backgr-fast-raht}


Pavez \textit{et al.}~\cite{raht} describe a re-formulation (henceforth P-RAHT) of the original RAHT algorithm~\cite{de2016compression} that exploits properties of the Morton code.
P-RAHT takes as input occupied voxels and their associated Morton codes.
It pre-computes, in a step called the \textit{RAHT} prelude, the constructs needed for the hierarchical RAHT computation, the sibling relationships and the node weights, since these are a function of voxel occupancy alone, not of the attribute.
Then, it re-uses these when encoding each attribute.


\parab{The RAHT prelude.}
This step computes three lists at every level $\ell \in \{1,\dots,3J\}$:
(i) an index list $I_\ell$ indicating those nodes in the binary tree at level $\ell$ that are either left siblings, or singletons (those with no occupied siblings); (the binary tree in \cref{fig:raht} shows only the nodes in these lists)
(ii) a weight list $W_\ell$ for those nodes (indicated by numbers inside nodes in \cref{fig:raht}, and
(iii) a flag list $F_\ell$ indicating which nodes are left siblings (not shown in the figure for brevity).

P-RAHT computes these lists bottom-up.
Consider a node $n$ at level $\ell$.
Its weight is the sum of the weights of its children at level $\ell+1$.
If either of its children is in $I_{\ell+1}$, node $n$ must be in $I_{\ell}$.
Finally, if $n$'s sibling at level $\ell$ is also in $I_{\ell}$, it is the left sibling if it has a lower Morton code.
For example, the rightmost node at the second level from the bottom has a weight of 2, because its children are both occupied.
For that reason as well, it belongs in the $I_\ell$ list for that level.
But, it does not belong in the corresponding $F_\ell$ list, since it is \textit{not} the left sibling at that level.

\parab{Computing RAHT Coefficients.}
Using these lists, P-RAHT computes the RAHT coefficients using another bottom-up pass for each attribute.
If a node $n$ at level $\ell$ is a left sibling (\textit{i.e.}, its flag is set in $F_\ell$), then it computes the low-frequency and high-frequency coefficients using~\cref{eq:raht-butterfly}.
Nodes \textbf{24} and \textbf{35} in \cref{fig:raht} are examples of left siblings.
It then passes the low-frequency coefficient to its parent and stores the high-frequency attribute in its right sibling.
For this, it uses the pre-computed weights in $W_\ell$.
If $n$ is a singleton, such as node \textbf{33} in \cref{fig:raht}, it simply passes its attribute to its parent.

\subsection{GPU-Accelerated RAHT}\label{sec:gpu-accelerated-raht}

  



  

\parab{Inputs and Outputs.}
\sysname{}'s GPU-accelerated RAHT takes as input occupied voxels, their Morton codes, and the attributes of the merged Gaussian in the voxel.
It outputs RAHT coefficients for all attributes.

\parab{Key Ideas.}
Our GPU-accelerated RAHT relies on three observations.
First, we observe that the basic RAHT algorithm is similar to the octree construction, encoding and decoding algorithms (\cref{sec:gpu-octree}).
RAHT proceeds bottom-up on a tree, performing one or more computations at some but not all nodes (\textit{e.g.}, only siblings), at each level in the tree.
The same is true for octree construction (\cref{sec:octree-encode-gpu}), for example, which performs computations only occupied nodes at each level.

This suggests that we can compute RAHT coefficients using the following design elements from those algorithms.
(a) Those algorithms use properties of Morton codes to mitigate compute stalls due to pointer chasing.
(b) They proceed sequentially level-by-level bottom-up or top-down, but within a level, they launch one GPU thread for every node in the tree involved in the computation.
(c) The set of nodes at a level involved in the computation is of variable length, so these algorithms employ PyTorch~\cite{pytorch} operations to convert fixed length vectors to variable length ones.
For example, the octree construction algorithm first creates a list of all occupied parents, then uses the \texttt{unique} operation to identify exactly the set of nodes at the next higher level involved in determining occupied nodes.

Our second observation is that P-RAHT is an ideal starting point for parallelization.
In particular, the pre-computed lists from the RAHT prelude greatly simplify applying the design elements described above.

Our third observation is that GPUs allow us to trivially \textit{parallelize RAHT coefficient computation across attributes}.
We can simply vectorize the attributes, and compute the coefficients in one bottom-up pass.

\parab{Computing RAHT coefficients on the GPU.}
\sysname{} computes RAHT coefficients bottom-up, exactly like P-RAHT.
Unlike P-RAHT, at each level, it launches multiple GPU threads.
Specifically, at level $\ell$, it launches a GPU kernel with one thread per entry in the list $I_\ell$.
For example, at the lowest level in \cref{fig:raht}, \sysname{} will launch \textbf{11} threads, and at the next level up, \textbf{7}.
Only threads whose flag $F_\ell$ is valid execute the $2\times2$ transform; each such thread reads indices of a sibling pair (current and next element in $I_\ell$), looks up their weights in $W_\ell$, and then applies Eq.~\ref{eq:raht-butterfly} to the attributes.

\parab{Parallelizing the RAHT Prelude.}
At each step $\ell$, \sysname maintains the current active list $I_\ell$ as a flat GPU tensor of indices.
We launch one GPU thread per element of $I_\ell$ to compute $W_\ell$ and $F_\ell$.
Each thread compares the Morton code of its current element with the next element (adjacent in Morton order) to decide whether they form a sibling pair.

After $F_\ell$ is computed, \sysname{} obtains the next active list $I_{\ell+1}$  by removing right siblings.
It implements this efficiently using boolean shift and masked select of tensors which are available in PyTorch (Alg.~\ref{algo:raht-prelude-gpu}).
The output of this is a compacted list containing exactly the set of nodes at $\ell+1$ that should run the RAHT coefficient computation at that level.

Especially for accelerating RAHT on a mobile GPU, we have found it essential to implement some of these steps using CUDA (
(\cref{sec:evaluation}),

\subsection{Quantization and Entropy Coding}\label{sec:quant-entr-coding}

After obtaining the RAHT coefficients for each attribute, \sysname{} \textit{quantizes} these coefficients.
Quantization trades off visual quality by reducing the number of bits required to represent a coefficient.
We quantize coefficients of different \gs{} attributes to different levels, similar to~\cite{mesongs,gpcc-adaptive}.
\sysname{} uses dead-zone quantization~\cite{deadzone}, which is simple to implement on a GPU, using one thread per coefficient, and vectorizing the attributes.

\sysname{} uses an entropy coder to efficiently serialize quantized coefficients into a bitstream.
We use the Run-Length Golomb-Rice (RLGR)~\cite{rlgr} entropy coder, which uses run-length coding for consecutive zeros, and Golomb-Rice coding for non-zero magnitudes using an adaptive parameter.
%
RLGR is inherently sequential, but \sysname{} parallelizes this by dividing the vector of coefficients, and entropy coding \textit{blocks} (\textit{e.g.}, 2K or 4K successive coefficients) using one GPU thread per block.
This ensures fast entropy coding at the expense of a slight loss in compression efficiency.

We have omitted a discussion of de-quantization and entropy decoding for brevity; \sysname{} also parallelizes this.
Thus, in \sysname{}, \textit{all steps} in octree and attribute encoding are GPU-accelerated.
\section{Improving \dgs{} Compression}\label{sec:klt}

In many \dgs{} videos, especially those that generate SH coefficients of degree 3, SH coefficients dominate the size of each \gs{} frame.
Some post-training compression techniques have realized this, and have developed specialized techniques to compress SH coefficients.
For example, MesonGS~\cite{mesongs} devises a vector quantization codebook for each frame, in order to compress SH coefficients.
This is extremely compute-intensive, so \sysname{} uses a different approach. 

In \gs{}, each SH function has 3 RGB color channels.
\sysname{} exploits the fact that RGB colors can be correlated~\cite{color_correlation}.
It first transforms RGB to YUV for each SH function.
In YUV, since Luma (Y) has more energy and it is more perceptually important than Chroma (UV), this transformation improves compression.
\sysname{} applies this RGB to YUV transformation, a simple matrix multiplication, which is fast on a GPU, to all SH  coefficients (both DC and AC).

\begin{figure}[t]
  \centering
  \begin{tikzpicture}
    \node (R) {\includegraphics[width=1.65in]{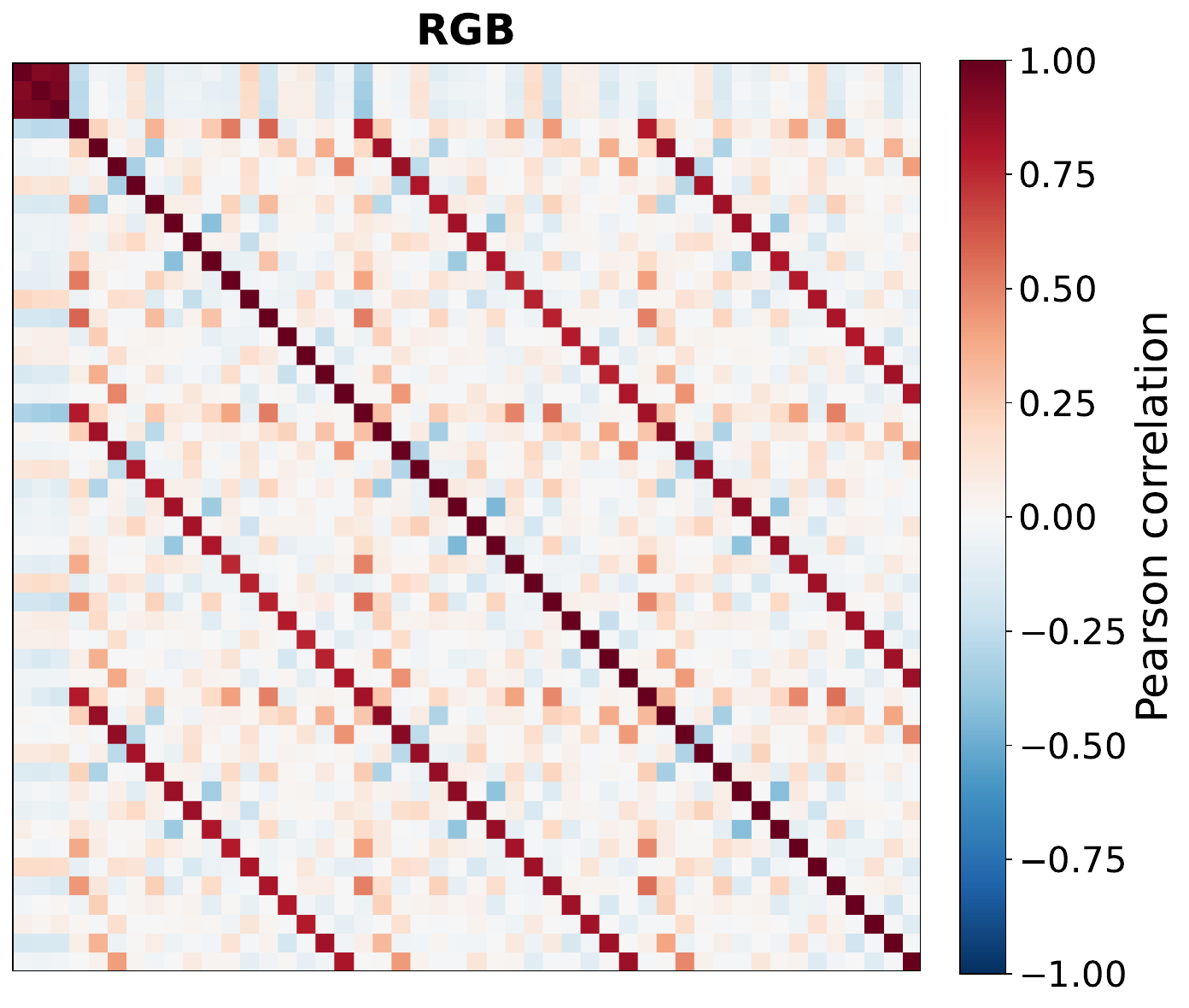}};
    \node[right=0mm of R] (K) {\includegraphics[width=1.65in]{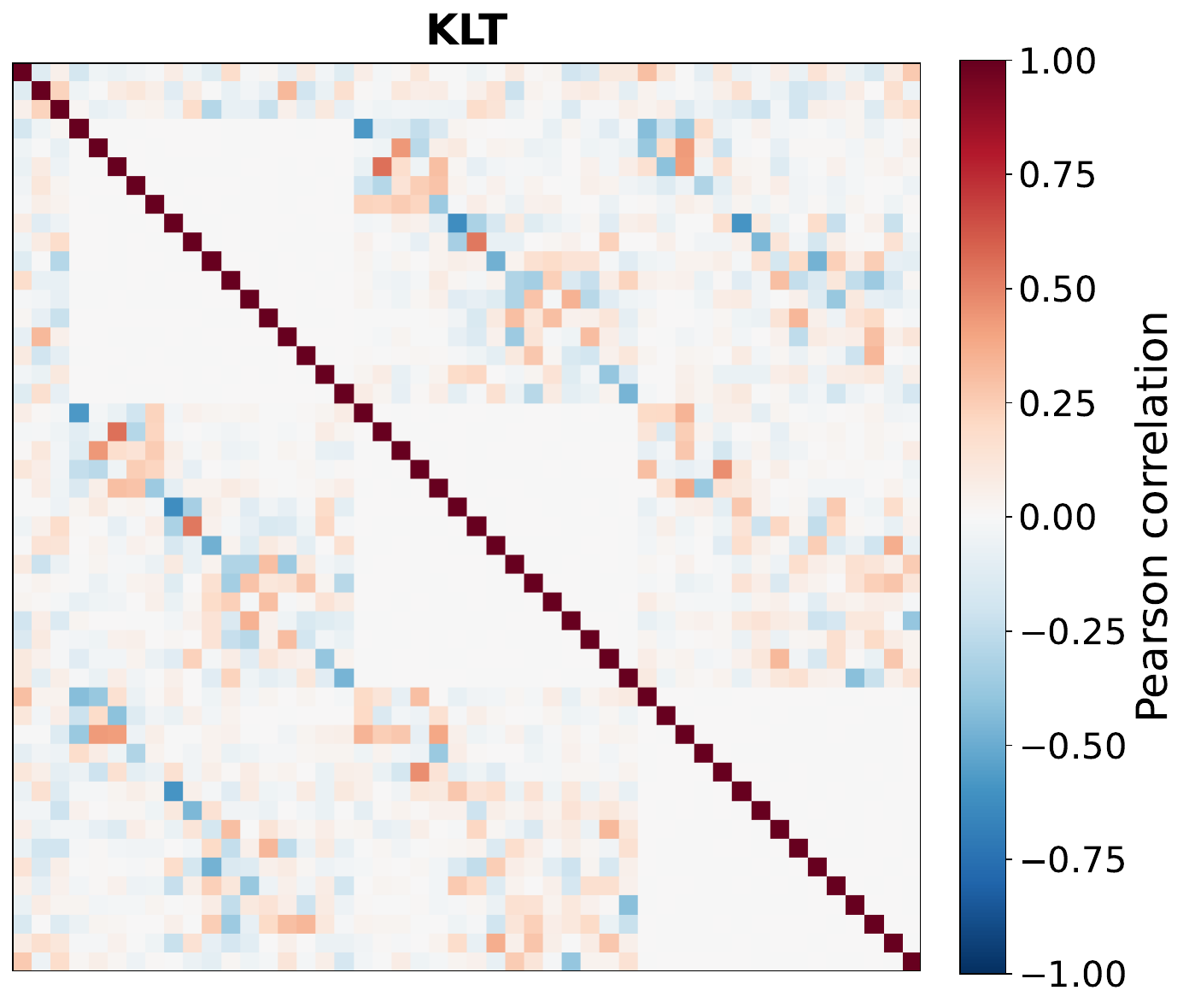}};
  \end{tikzpicture}  
  \caption{\label{fig:rate-dist-example} Heatmaps that plot Pearson correlation across SHs in RGB domain (left) and after YUV conversion and KLT (right).}
\end{figure}

While RGB to YUV conversion is common in image and video compression, because \gs{} represents color with high dimensional SH coefficients, there are additional opportunities to remove redundancy. After YUV conversion, \sysname{} applies a decorrelating transform, Karhunen-Loève transform (KLT~\cite{KLT}), \textit{only to the AC coefficients} (\textit{i.e.}, to SH coefficients of degree greater than 0). More specifically, we apply three different KLTs, one  to each channel in YUV color space. 
The KLT effectively concentrates signal energy on the first few coefficients.
\cref{fig:rate-dist-example} illustrates this.
On the left is a heatmap of the Pearson correlation coefficients for the original RGB SH coefficients.
On the right is the corresponding heatmap after applying  RGB to YUV conversion followed by KLT; notice how the figure on the right has significant white-space (white corresponds to zero correlation).
%

By decorrelating color across AC SH coefficients, KLT reduces redundancy, concentrates energy, and  thereby improves compression.
We apply KLT \textit{before} RAHT (\cref{fig:architecture}).
KLT requires a matrix multiplication, which we implement on the GPU with minimal overhead.

\section{Evaluation}\label{sec:evaluation}

In this section, we compare \sysname{} against the state of the art, using two popular \dgs{} data sets.

\subsection{Methodology}\label{sec:methodology}


\parab{Implementation.}
We have implemented \sysname{} in Python.
It uses custom CUDA kernels for the octree encoder/decoder (\cref{sec:gpu-octree}), RAHT (\cref{sec:gpu-raht}), and RLGR  (\cref{sec:quant-entr-coding}).
It compresses position using the ANS entropy coder from NVIDIA's nvCOMP library~\cite{nvidia_nvcomp} and implements KLT decorrelation and quantization in PyTorch~\cite{pytorch}; these run on the GPU.
Most of the data and intermediate tensors use PyTorch.
%
%
%

\parab{Hardware.}
Most experiments run on a PC equipped with an NVIDIA RTX 4500 Ada Generation GPU (24~GB VRAM), 128~GB system RAM, and Intel(R) Xeon(R) w7-2475X with 20 physical cores (40 threads with hyperthreading).
Mobile decoding runs on an NVIDIA Jetson.

\parab{Baselines.}
We compare \sysname{} against four post-training compression baselines:
\begin{description}
  \item[\vthree{}-2D~\cite{v3volumetric}]
  maps Gaussian attributes to 2-D image stacks, one per attribute channel, and compresses the resulting images with a H.264 codec using \texttt{x264} from FFmpeg~\cite{ffmpeg}.
  It compresses each channel of attributes into a single MP4 stream, where frames for that attribute form a group of pictures (GOP), and a single quantization parameter (QP) controls the quality-size tradeoff.
  It splits the position of Gaussians into MSB and LSB, where QP for MSB is 0 (lossless) and QP for LSB is the QP passed to the encoder (default: 25).
  \vthree{}-2D caps QPs for rotation and scale attributes to 22, and quantizes all other attributes (position LSB, color, opacity) with the same QP (default: 25).
  It compresses each attribute channel using one call of the H.264 encoder, resulting in 62 MP4 files per GOP (default: 20).

  \item[MesonGS~\cite{mesongs}] is a partly GPU-accelerated codec that uses an octree for geometry, RAHT and block quantization for attributes, vector quantization for higher-degree SH coefficients, and LZ77 for entropy coding. 
  We use the authors' default parameters.

  \item[G-PCC~\cite{gpcc}] Wang \textit{et al.}~\cite{gpcc-adaptive} use G-PCC to compress static \gs, and adapt octree depth for voxelization to preserve quality.
  This requires fine-tuning of Gaussians after voxelization to reach high quality.
  To obtain a baseline that directly extends G-PCC to \dgs{} videos, we implement a simplified version of \cite{gpcc-adaptive} (without adaptive voxelization and fine-tuning).
  The approach uses RAHT for color and opacity (lossy, depending on QP) and hierarchical neighborhood prediction for rotation and scale (lossless).

  \item[LTS-Draco~\cite{LTS,mga}] 
  uses modified Draco to independently compress each \gs{} frame on the CPU. 
  The number of quantization bits per attribute is configurable (default: 16 bits for all attributes, compression level 10).
\end{description}
In each case, we use the implementations provided by the authors.
Moreover, these baselines use qualitatively different compression techniques for \dgs{}: 2D codecs~\cite{v3volumetric}, Draco~\cite{LTS}, and G-PCC~\cite{mesongs,gpcc-adaptive}.

\parab{Datasets.}
We evaluate on two widely-used \dgs{} video datasets spanning diverse content types:
\begin{description}
  \item[HiFi4G~(7 sequences)] contains person-centric captures of actors performing various actions, with $200$ frames per sequence. We use \gs{} models trained using \vthree~\cite{v3volumetric} with SH coefficient degrees $0..3$ (${\sim}120$K--$300$K Gaussians per frame).

  \item[Neural 3D Video (N3DV)~(6 sequences)] contains full-scene captures of indoor table-top activities with day and night lighting, each with~300 frames. We use models trained with QUEEN~\cite{queen} (without its in-training compression) with SH degrees $0..2$ (${\sim}150$K--$400$K Gaussians per frame).
\end{description}

\parab{Metrics.}
For each approach, we report measures of:
(1) \textit{visual quality} based on PSNR computed on rendered test views of the decompressed frames against the uncompressed ground truth (we use the test views used by \vthree and QUEEN);
(2) \textit{per-frame encode} and \textit{decode latency} in milliseconds (ms); and
(3) \textit{compression efficiency} in terms of \gs{} frame size after compression.
We describe the precise metrics below.

%

\subsection{Baseline Comparison}\label{sec:baseline-comparisons}




  

\parab{Methodology.}
This section compares \sysname{} against all baselines at a single operating point.
For competing approaches, we select the default parameters for compression, such as QP, octree depth, bit-depth, \etc, that we either found in the corresponding paper or source code.
For \sysname{}, we use parameters that give us the same quality (PSNR) on the first frame of a sequence as the default setting for each approach.
For instance, if, on the first frame of sequence $S$, \vthree{}-2D has a PSNR of $p$, we find a parameter setting for \sysname{} (using the technique described in \cref{sec:rd-curves}) whose PSNR for the first frame on $S$ is close to $p$.
This results in a pair-wise comparison between \sysname{} and each baseline.

We use these settings to encode, decode, and estimate PSNR and compression sizes for every 10th frame\footnote{For \vthree{}-2D, we compute the size of a frame as the average size of the GOP that it belongs to.} of every sequence in the two datasets listed above.
We do this primarily because some of the baselines~\cite{mesongs,gpcc-adaptive} have exceedingly high encode/decode latencies, rendering an exhaustive evaluation intractable.
Then, we compute, for each sequence: (a) the average per-frame encode latency, (b) the average per-frame decode latency, (c) the average per-frame PSNR difference between \sysname{} and the baseline (the \dpsnr{}), (d) average per-frame size ratio between \sysname{} and the baseline (the \textit{relative compression ratio}, or \rcr{}).

\begin{table}[t]
  \centering
  \footnotesize
  \begin{tabular}{l|rr|rr}
    \hline
    & \multicolumn{2}{c|}{\textbf{N3DV (QUEEN)}} & \multicolumn{2}{c}{\textbf{HiFi4G (\vthree)}} \\
    \textbf{Pipeline} & Enc.\ (ms) & Dec.\ (ms) & Enc.\ (ms) & Dec.\ (ms) \\
    \hline\hline
    \textbf{\sysname{}} & \textbf{23} & \textbf{21} & \textbf{18} & \textbf{14} \\
    \vthree{}-2D & 370 & 290 & 1098 & 341 \\
    LTS-Draco   & 400 & 158 & 156 & 91 \\
    G-PCC        & 8804 & 6928 & 6101 & 3650 \\
    MesonGS     & 28996 & 1096 & 144953 & 535 \\
    \hline
  \end{tabular}
  \caption{Per-frame encode and decode latency (ms).}
  \label{tab:baseline-latency}
\end{table}

\subsubsection{Encode-Decode Latency}\label{sec:latency}

\cref{tab:baseline-latency} summarizes per-frame encode and decode times aggregated across all sequences within each dataset.
We do this because, within a dataset (\textit{e.g.}, HiFi4G), there is little variability in encode and decode times across sequences, since all sequences are person-centric.
\cref{app:coding-latency} contains a more detailed breakdown.

\sysname{} achieves ${\sim}23$~ms encode and ${\sim}21$~ms decode on N3DV, and ${\sim}18$~ms encode and ${\sim}14$~ms decode on HiFi4G---well under the 33~ms budget for 30 frames-per-second (fps).

LTS-Draco, the next fastest baseline, is \textbf{9--17$\boldsymbol\times$ slower than \sysname{} for encoding} and \textbf{7--8$\boldsymbol\times$ slower for decoding}.
Draco encoding and decoding could potentially be made faster using coarse-grained CPU parallelism~\cite{meshreduce,vivo,metastream}, but this trades off compressibility.

Though 2D video codecs are fast, \vthree{}-2D's several hundred milliseconds to encode decode, due to the overhead of calling video encoder sequentially for each channel\footnote{\vthree{}-2D encodes and decode the full scene N3DV dataset faster than the HiFi4G person-centric dataset, because, on the former dataset, QUEEN does not produce degree-3 SHs.};
It is \textbf{16--61$\boldsymbol\times$ slower than \sysname{} for encoding} and \textbf{14--24$\boldsymbol\times$ slower for decoding}.
\vthree{}-2D could offload H.264 encoding to an NVIDIA GPU using \texttt{nvenc/nvdec}~\cite{nvenc-sdk}; however, \texttt{nvenc} allows only 8 parallel encoders on a desktop-class GPU~\cite{nvenc-limit}, and \vthree{}-2D needs to encode 62 videos, so it is unlikely to achieve full-frame rate performance with offloading.

G-PCC has the next-highest encode and decode latencies; on both data sets, its encode and decode latencies are at least \textbf{260}$\times$ larger than \sysname{}'s.
Although \sysname{}'s design borrows heavily from G-PCC elements, its GPU-acceleration results in a significant performance difference.

MesonGS is the slowest baseline.
It also borrows heavily from G-PCC, and uses some GPU acceleration, but takes several seconds or minutes to encode and decode each \gs frame, mostly because of the time to learn the vector quantization codebook for AC coefficients~\cite{mesongs}.
%
Its decoding time is dominated by RAHT decoding which takes up to $200$--$600$~ms. 

Overall, \sysname{} can encode and decode \textit{one or two orders of magnitude} faster than the baselines on our datasets.



\begin{table}
    \centering
    \scalebox{0.55}{
        \begin{tabular}{llccccccc}
        \toprule
        Sequence & \multicolumn{2}{c}{\vthree{}-2D} & \multicolumn{2}{c}{MesonGS} & \multicolumn{2}{c}{LTS-Draco} & \multicolumn{2}{c}{G-PCC} \\
                 & \dpsnr{} (dB) & \rcr{} & \dpsnr{} (dB) & \rcr{} & \dpsnr{} (dB) & \rcr{} & \dpsnr{} (dB) & \rcr{} \\
        \midrule
        Actor1 & 0.02 & 0.52 & -0.02 & 2.00 & -0.04 & 3.01 & -0.03 & 4.91 \\
        Actor2 & 0.03 & 0.31 & -0.03 & 2.11 & -0.04 & 3.05 & -0.01 & 5.08 \\
        Actor3 & -0.01 & 0.64 & 0.01 & 1.74 & -0.02 & 3.18 & 0.05 & 4.52 \\
        Actor4 & -0.05 & 0.56 & -0.02 & 1.77 & -0.07 & 2.84 & 0.01 & 4.98 \\
        Actor5 & 0.07 & 0.53 & 0.05 & 1.60 & -0.07 & 2.91 & 0.01 & 5.06 \\
        Actor6 & -0.07 & 0.81 & -0.18 & 2.07 & -0.09 & 2.87 & 0.01 & 4.66 \\
        Actor7 & 0.02 & 0.50 & -0.08 & 2.05 & -0.04 & 3.01 & 0.00 & 4.77 \\
        \hline
        Mean & 0.00 & 0.55 & -0.04 & 1.91 & -0.05 & 2.98 & 0.01 & 4.85 \\
        \midrule
        cook\_spinach & 0.05 & 0.48 & 0.04 & 1.68 & -0.19 & 4.27 & -0.02 & 6.46 \\
        coffee\_martini & 0.39 & 0.67 & -0.06 & 1.52 & 0.01 & 6.38 & -0.04 & 3.26 \\
        cut\_roasted\_beef & 0.06 & 0.35 & -0.01 & 1.34 & -0.08 & 3.72 & -0.19 & 6.60 \\
        flame\_salmon & 0.74 & 0.61 & 0.09 & 1.44 & -0.02 & 8.17 & -0.01 & 3.40 \\
        flame\_steak & -0.08 & 0.38 & -0.03 & 1.30 & -0.16 & 4.04 & 0.39 & 7.35 \\
        sear\_steak & -0.11 & 0.31 & -0.01 & 1.45 & -0.17 & 4.04 & 0.20 & 5.54 \\
        \hline
        Mean & 0.18 & 0.47 & 0.00 & 1.46 & -0.10 & 5.10 & 0.06 & 5.44 \\
        \bottomrule
        \end{tabular}
    }
    \caption{Per-frame PSNR difference and compression ratio difference between \sysname{} and the baseline. For every metric, higher is better for \sysname{}.
    }
    \label{tab:baseline-psnr-comp}
  \end{table}

\subsubsection{Quality and Compression Performance}\label{sec:qual-compr-perf}

\cref{tab:baseline-psnr-comp} shows the \dpsnr{} and \rcr{} performance for each sequence from both datasets, across all baselines.
To understand the results, recall that:
(a) \dpsnr{} captures the PSNR difference between \sysname{} and a baseline, and \rcr{} the relative frame sizes after compression between the baseline and \sysname{};
(b) when we do this experiment, we find a setting for \sysname{} which has approximately the same PSNR on the first frame as the baseline's default setting; and
(c) the values in the table average these quantities across measured frames.
Finally, a positive value of \dpsnr{} indicates that \sysname{} has better quality, and a value $> 1$ for \rcr{} indicates that \sysname{} has better compression performance.
Given this, we expect \dpsnr{} to be close to zero for all sequences, for all approaches.
This is because the sequences do not have significant scene changes, so a parameter setting from the first frame is likely to give mostly similar PSNR values for all other frames, for all baselines.

\parab{\vthree{}-2D.}
For this approach, \dpsnr{} is near zero for all sequences in HiFi4G.
However, even though we roughly equalized PSNR, \vthree{}-2D has nearly 0.4 dB lower PSNR in the \textit{coffee\_martini} sequence, and a 0.74 db lower PSNR in the \textit{flame\_salmon} sequence in the N3DV data set.
\vthree{}-2D has evaluated their approach on person-centric datasets, not on the larger scenes in this data set.
We discuss this in greater detail in \cref{sec:rd-curves}, where we show that this approach fails to achieve high quality on some N3DV sequences.

However, \vthree{}-2D has better compression performance than \sysname{}.
On HiFi4G, for example, \sysname{}'s frame sizes can be \textbf{1.2--3}$\times$ larger than \vthree{}-2D.
On N3DV, \sysname{}'s frame sizes can be \textbf{1.2--2.6}$\times$ larger than \vthree{}-2D.
On average, for both datasets, \vthree{}-2D's frame size is about half that of \sysname{}.
%
By using a 2D codec, \vthree{}-2D can exploit inter-frame compression techniques in those codecs.
In contrast, all other techniques, including \sysname{} only use intra-frame compression, so have lower compression performance.

\parab{MesonGS.}
For this approach, \dpsnr{} is uniformly near zero for all sequences, as expected.
However, it is less compression-efficient than \sysname{}.
On the HiFi4G, data set, MesonGS's frame size is nearly 2$\times$ that of \sysname{}, and on N3DV it is 1.46$\times$.
MesonGS uses similar algorithms as ours, but with two important differences.
They use vector quantization for SH coefficients\footnote{MesonGS~\cite{mesongs} was designed to encode static scenes and we have used it to encode \dgs{}. Most video codecs would only retrain the vector quantization codebook every few frames, so our approach slightly over-estimates MesonGS frame sizes}, but we apply KLT (\cref{sec:klt}) and then quantize RAHT coefficients.
They also use LZ77 for encoding attributes and positions, we use ANS and RLGR entropy encoders.
%
We conjecture that these differences contribute to MesonGS's lower compression performance and to its higher encode-decode latencies (\cref{tab:baseline-latency}).

\parab{LTS-Draco.}
For this approach, \dpsnr{} is near zero, except for a few sequences in N3DV, where LTS-Draco exhibits a \textbf{0.16--0.19} dB PSNR drop.
In conducting this experiment, we sought settings for \sysname{} whose PNSR matched that of LTS-Draco, but this is not always possible, so we selected the configuration that provided the closest match.
This is likely the reason for the discrepancy.
LTS-Draco, however, has much worse compression performance; its frame sizes are nearly 3$\times$ \sysname{}'s for HiFi4G and 5$\times$ for N3DV.

\parab{G-PCC.}
Of all the approaches, G-PCC has the highest \rcr{}; its compression performance is 3.4$\times$ to 7.35$\times$ worse than \sysname{}.
Although \sysname{} uses similar components, the adaptation~\cite{gpcc-adaptive} of G-PCC to \dgs{} that we use compresses SH AC coefficients far less aggressively than \sysname{}.
That adaptation also applies \textit{lossless compression} to scales and rotations (that is, it does not quantize coefficients before entropy coding), while \sysname{} applies \textit{lossy compression} by quantizing those attributes.
These two factors contribute to the significant differences.

\subsection{Rate-Distortion Curves}\label{sec:rd-curves}



\begin{figure*}[t]
  \begin{minipage}[t]{0.33\textwidth}
    \centering
    \includegraphics[width=0.9\columnwidth]{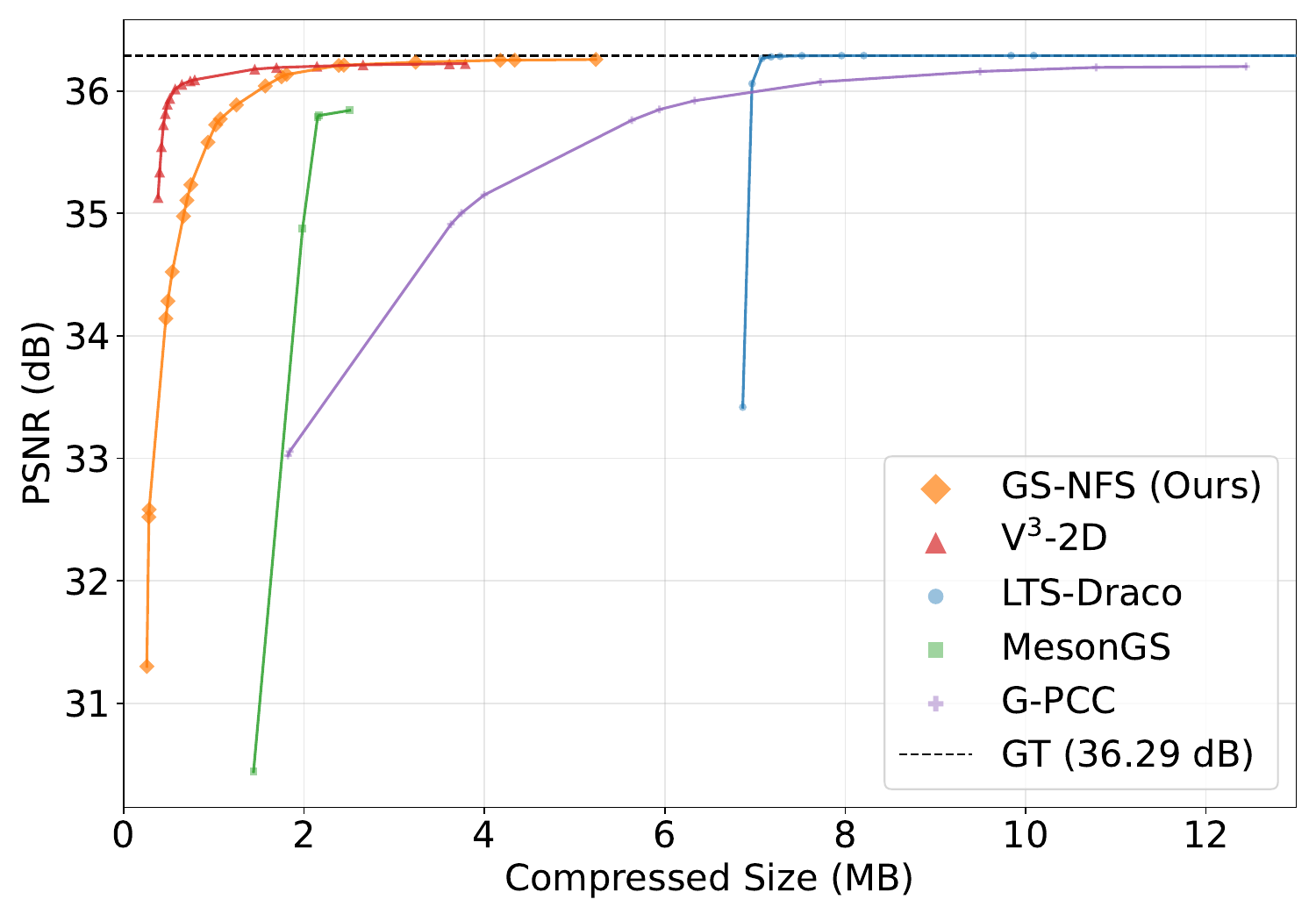}
    \label{fig:rd-act1}
  \end{minipage}
  \begin{minipage}[t]{0.33\textwidth}
    \centering
    \includegraphics[width=0.9\columnwidth]{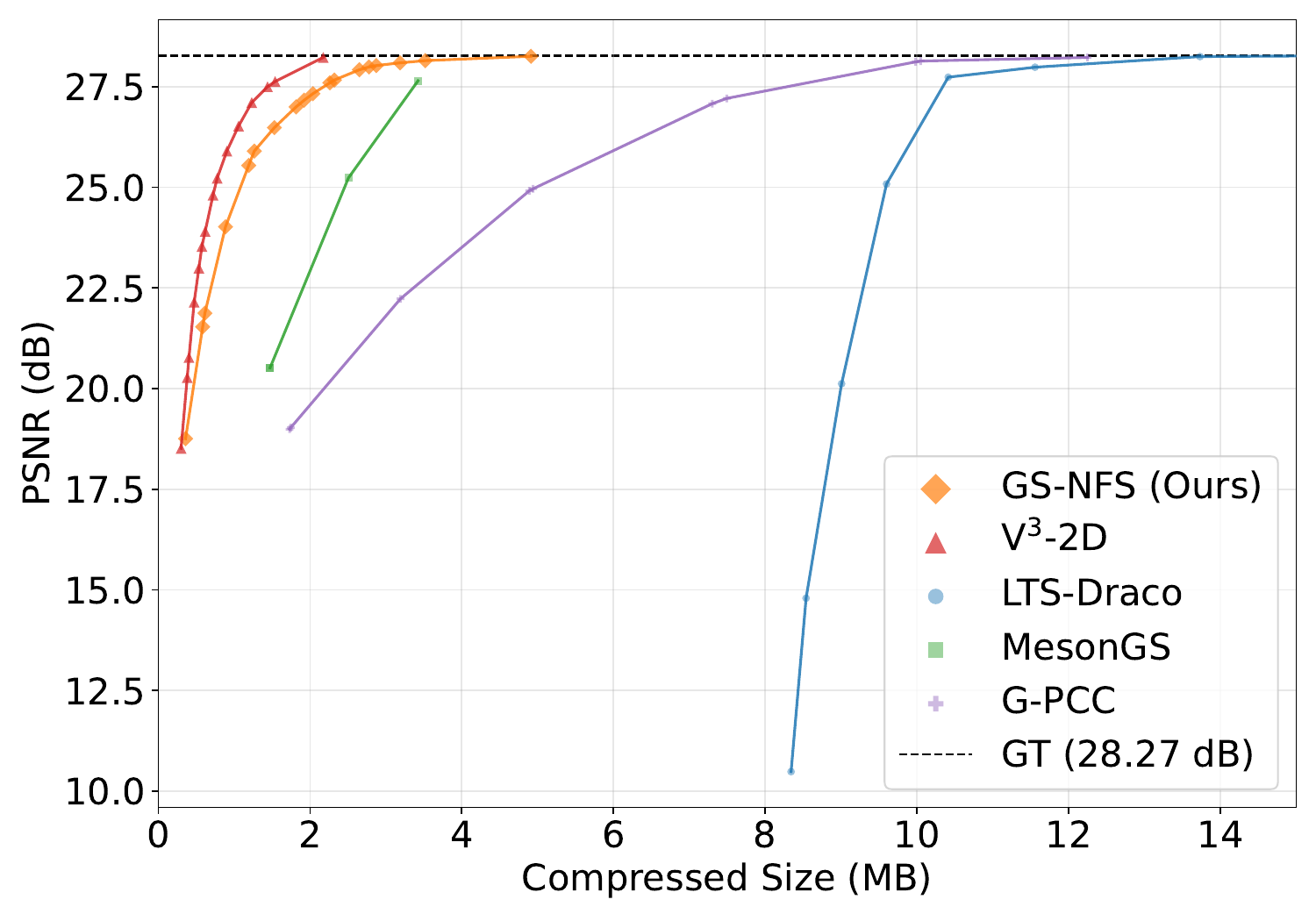}
    \label{fig:rd-salmon}
  \end{minipage}
  \begin{minipage}[t]{0.33\textwidth}
    \centering
    \includegraphics[width=0.9\columnwidth]{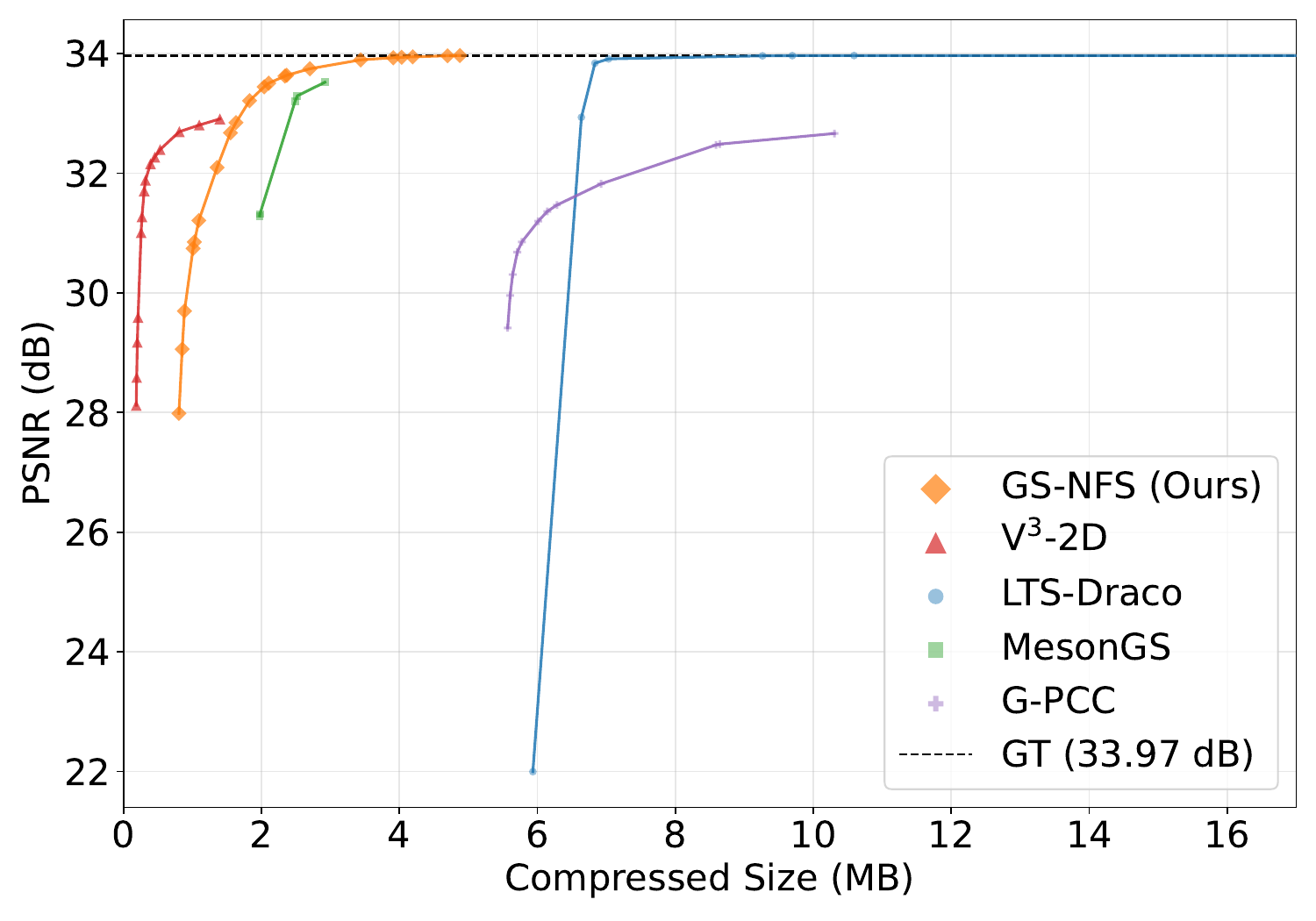}
    \label{fig:rd-steak}
  \end{minipage}
  \caption{R-D curves for (Left) Actor1 from HiFi4G, (Middle) flame\_salmon from N3DV, and (Right) sear\_steak from N3DV.}
  \vspace{-10pt}
  \label{fig:rd-curves}
\end{figure*}

To understand quality and compression trade-offs of different designs, we can also use a \textit{rate-distortion} curve or \textit{R-D curve}.
Such a curve plots \textit{rate} on the x-axis, \textit{quality} or distortion on the y-axis, and the curve represents the best quality one can obtain for a given rate.
To obtain this curve, practitioners explore a large space of codec parameters, obtain the rate and quality for each parameter, and plot these as points on the R-D plot.
The R-D curve is the \textit{convex hull} of those points.

R-D curves can also help encode videos at a given bitrate $B$.
To do this, we find the point on the convex hull whose rate is closest to $B$, and use its configuration to encode at the target bitrate.
We also used R-D curves to find the \sysname{} configuration with the nearest PSNR in \cref{sec:qual-compr-perf}.

\parab{Methodology.}
We computed R-D curves for the first frame of \textit{all} sequences, for all our baselines.
To do this, we sweep the space of parameters of each baseline and \sysname{}, plot rate and distortion and compute the convex hull as discussed above.
\cref{app:generating-r-d} discusses the parameters we used for each baseline and for \sysname{}.
For some approaches like MesonGS and G-PCC, we could not explore all parameter combinations because of their high encoding latency, and exploring the entire space would have taken hours or days.
In contrast, for \sysname{}, we were able to evaluate 1000 combinations in about 12 minutes.



\parab{R--D curves.}
\cref{fig:rd-curves} shows the R--D curves for one HiFi4G sequence (\textit{Actor1}) and two N3DV sequences (\textit{flame\_salmon\_1} and \textit{sear\_steak}).
\cref{fig:rd-baselines-hifi4g} and \cref{fig:rd-baselines-n3dv} depict the R-D curves for the remaining sequences for the two datasets respectively.
If the R-D curve for approach $A$ is entirely to the left of the curve for $B$, for a given sequence, we say that $A$ dominates $B$.
If the two curves intersect, then one is better in some bitrate regimes, and worse in others.

Consider the R-D curves for \textit{Actor1}.
Clearly, \sysname{} dominates MesonGS, Draco-LTS, and G-PCC.
However, \vthree{}-2D dominates \sysname{}: it can, for a given quality, encode at a lower rate than \sysname{}.
This explains why its compression performance was better in our baseline comparisons (\cref{sec:qual-compr-perf}), and is likely due to its use of 2D codecs' ability to do inter-frame compression.

The same is true for \textit{frame\_salmon}, except for this full-scene sequence, \vthree{}-2D's R-D curve is closer to, but still dominates \sysname{}'s.
For this sequence, both these approaches dominate MesonGS, Draco-LTS, and G-PCC.

On the other hand, for \textit{sear\_steak}, \vthree{}-2D does \textit{not} dominate \sysname{}.
In fact, for this sequence, and for \textit{three other sequences} in N3DV out of a total of six (cook\_spinach, cut\_roasted\_beef, and flame\_steak, \cref{fig:rd-baselines-n3dv}), \vthree{}-2D's R-D curve shows that \textit{there is no parameter setting for which its quality reaches close to the ground-truth PSNR} (we explored QP values from 1 to 40, where 1 is the best quality).

In each case, \sysname{} has a parameter setting that can provide \textbf{0.5--2 dB} higher PSNR for the same rate.
Moreover, for all of these sequences, \sysname{} has parameter settings that can reach ground-truth (GT) quality.
Put another way, if \sysname{} were used to encode videos for DASH-based \dgs{} streaming, for these sequences, it would provide much better quality at higher bandwidth-availability.

To our knowledge, \vthree{}-2D has only been evaluated on person-centric datasets and not on full-scene ones.
Now, \vthree{}-2D must map 3D Gaussian positions to a 2D image.
%
%
%
%
There are many possible 3D-to-2D mappings; \vthree{}-2D applies a fixed criterion to determine where an attribute corresponding to a Gaussian in 3D should be located in the 2D.
Consider two consecutive frames in 3D; the 3D-to-2D mapping algorithm is applied independently to each frame, which means that even if motion in the 3D scene is smooth, Gaussian attributes close to each other on the first 2D frame may not be close to each other in the second 2D frame. 
As a sequence, because the location of the same 3D attributes may change from one 2D frame to another, motion-compensated prediction in the 2D frames may not work well, potentially resulting in a higher rate than an intra-only approach. 
We conjecture that, for this reason, \vthree{}-2D does not perform well in some N3DV sequences, where inter-frame coding becomes inefficient.
On the other hand, \sysname{} does not exhibit this degradation because it uses a 3D codec.
%



\subsection{Mobile Performance}\label{sec:mobile-performance}

\sysname{} supports \dgs{} playback on resource-constrained devices.
To demonstrate this, we evaluate \sysname{}'s decode performance on an NVIDIA Jetson Orin, a mobile GPU with compute comparable to those in in AR/VR headsets and edge devices~\cite{nvidia_jetson_orin}.
%
\cref{tab:jetson-decode} shows end-to-end decode latency on Jetson Orin.
We observed that many \gs{} mobile implementations can support 30 fps \gs{} rendering for Gaussians with only SH DC coefficients (SH0), but not at higher SH degrees~\cite{v3volumetric}.
At SH0, \sysname{} achieves 40--59~ms across all sequences, allowing about $17$--$25$~fps decoding.
Latency scales \emph{sub-linearly} with channel count: going from SH0 to SH3 (5.1$\times$ more channels) increases average decode latency across all sequences by ${\sim2}\times$.
At higher SH degrees, scene complexity dominates for full-scene content (more Gaussians): flame\_salmon at SH2 (35~ch) is $2\times$ slower than Actor1 at SH3 (56~ch), despite having fewer channels.

\begin{table}[t]
  \centering
  \footnotesize
  \begin{tabular}{l|l|r|r}
    \hline
    \textbf{Sequence} & \textbf{SH degree} & \textbf{\# channels} & \textbf{Decode latency (ms)}\\
    \hline\hline
    Actor1         & 0 & 11 & 40 \\
    Actor1         & 3 & 56 & 57 \\
    flame\_salmon  & 0 & 11 & 59 \\
    flame\_salmon  & 2 & 35 & 120 \\
    sear\_steak    & 0 & 11 & 49 \\
    sear\_steak    & 2 & 35 & 113 \\
    \hline
  \end{tabular}
  \caption{Decode latency on Jetson Orin, averaged across frames.}
  \label{tab:jetson-decode}
\end{table}

\subsection{Novel \sysname{} Use Cases}\label{sec:other-capabilities}


So far, we have discussed using \sysname{} to compress stored \dgs{} videos.
In this section, we demonstrate that \sysname{} can \textbf{compress, on-the-fly, live \dgs{} videos} generated by feed-forward Gaussian generators and \textbf{point-clouds} as well, both at full frame-rate.
%
%
\cref{app:compr-stat-scen} also demonstrates \sysname{}'s ability  to quickly compress large static \gs{} scenes.

\parab{Live \dgs{} Compression.}
GPS-Gaussian~\cite{zheng2024gpsgaussian} generates Gaussian at ${\sim}25$~fps (${\sim}270$K Gaussians per frame) using captures from 2 views of a person.
We adapted \sysname{} to compress these frames on-the-fly.
Specifically, our approach repeatedly takes 2 views of a person per frame (pre-loaded from disk), generates a \gs{} frame using GPS-Gaussian, compresses the frame using \sysname{}, and finally decompresses the frame to reconstruct the \gs frame.
This pipeline can achieve \textbf{22~fps} and an end-to-end latency (from capture to after decode) of \textbf{75~ms}.
While \sysname's encoder and decoder takes only $15$~ms each, the GPS-Gaussian generator takes ${\sim}43$~ms, and is the bottleneck in this pipeline.

\parab{Point cloud compression.}
Since \sysname{} builds upon G-PCC, it can compress point clouds (\gs with only DC attribute).
So, we evaluate its performance on popular point cloud-based volumetric video datasets.
We use six sequences from two datasets: four single person sequences from 8i~\cite{8i-ptcl} (${\sim}700$K--1M points/frame; ${\sim}17$--$26$~MB) and 2 multi-person full-scene sequences from CMU Panoptic~\cite{cmu-panoptic} (${\sim}700$K--850K points/frame; ${\sim}17$--$21$~MB).
Point clouds are voxelized at octree depth $J\!=\!10$ and \sysname{}'s codec uses a quantization step $0.1$ for color attributes.
%
\sysname{} can encode each frame in $12$--$14$~ms and decode in $9$--$12$~ms while achieving \textbf{lossless geometry} and color Y-PSNR of $31$--$37$~dB. 
It can compress 8i sequences to $270$--$370$~KB and CMU Panoptic sequences to $660$--$780$~KB, \textbf{26--64$\times$} smaller than the original size.

\subsection{Ablations}\label{sec:ablations}

%

\parab{KLT color decorrelation.}
\sysname{} decorrelates SH coefficients using a YUV transform followed by a per-channel KLT (\cref{sec:klt}).
%
%
\cref{tab:ablation-klt} shows the impact on compressed size.
RGB inflates the attribute bitstream by \textbf{43\%} compared to raw RGB, while YUV is larger than KLT by \textbf{36\%} for flame\_salmon.
For others, the reductions are smaller, but still significant.
The additional latency for computing and applying the KLT matrix is less than 1~ms per frame, a negligible overhead.
%

\parab{CUDA vs.\ PyTorch RAHT decode on Jetson.}
RAHT takes about ${\sim}50$\% of the per frame decode latency, particularly on Jetson Orin.
An efficient implementation of RAHT is crucial for mobile decode.
\sysname{}'s custom CUDA RAHT kernels achieve $2.0$--$3.9\times$ speedup over a PyTorch baseline on Jetson, with larger gains for lower SH degrees (\cref{tab:jetson-raht-decode}).
We see a reduction of about $4$--$5\times$ for SH0 and $2$--$4\times$ for higher SH degrees for RAHT decode, which is the main bottleneck for mobile decode; prelude has ${\sim}2\times$ of speedup, since it doesn't depend on the number of channels.
These gains also reduce encode latency, since both forward and inverse RAHT perform the same work.

\parab{Other Ablations.}
\cref{app:other-ablations} describes the cost of CPU octree coding (\cref{sec:gpu-octree}), and explores RLGR parameter sensitivity of (\cref{sec:quant-entr-coding}).



%

%


\section{Related Work}\label{sec:related-work}
%

\parab{During-training \gs compression.}
Recent research has focused on integrating compression directly into an optimization loop within training to enhance model sparsity.
Compact3DGS~\cite{compact3dgs} identifies prunes redundant Gaussians (as do~\cite{recongs2025, scaffoldgs}) and replaces spherical harmonics with hash-based grids.
CompGS~\cite{compgs} utilizes quantization-aware training and opacity regularization.
4D-GS~\cite{4dgs} trains an MLP to predict motion across frames, while QUEEN~\cite{queen} learns quantized attribute residuals and uses a gating module to sparsify positions. 
%
%
LapisGS~\cite{shi2025lapisgs} trains \gs{} in a layered progressive representation.
Relative to \sysname{}, these methods incur significantly higher overhead for compression.

\parab{Post-training \gs compression.}
%
MesonGS~\cite{mesongs} prunes insignificant Gaussians and then applies octree and RAHT.
Some schemes~\cite{gauspcgc, entropygs} use learning based methods to find the compression parameters for position and attributes.
Others~\cite{LTS, gpcc-adaptive} use point cloud compression techniques to compress trained \gs{}.
%
%
To address this, \vthree{}~\cite{v3volumetric} encodes \dgs{} in multiple 2D videos.
\sysname{} encodes and decodes much faster than these techniques.
%

\section{Conclusions}\label{sec:conclusions}

\sysname{} is a frame-rate GPU-accelerated encoder and decoder for \dgs{} videos.
To our knowledge, it is the first to demonstrate frame-rate encoding and decoding performance not just for \dgs{}, but also for point clouds.
It achieves this performance using novel parallelization algorithms for position and attributes.
Its compression performance is also better than many state-of-the-art approaches due to a novel de-correlation approach.
Future work can develop rate-adaptive codecs that can dynamically adapt compression levels in response to bandwidth changes, using \sysname{}'s ability to quickly encode and decode frames.

\bibliographystyle{ACM-Reference-Format}
\bibliography{references.bib}

\newpage
\appendix
\renewcommand\thefigure{A.\arabic{figure}}
\renewcommand\thesection{\Alph{section}}
\renewcommand{\thealgocf}{A.\arabic{algocf}}
\renewcommand\thetable{A.\arabic{table}}
\setcounter{section}{0}
\setcounter{algocf}{0}
\setcounter{figure}{0}
\setcounter{table}{0}

\section{Coding Latency}\label{app:coding-latency}

In \cref{tab:latency_per_sequence}, we report the actual encoding and decoding latencies for all sequences in the HiFi4G and N3DV datasets.

\begin{table*}[t]
  \centering
  \footnotesize
  \caption{Encode and decode latency (ms) for video sequences.}
  \label{tab:latency_per_sequence}
  \begin{tabular}{lrrrrrrrrrr}
  \toprule
  Sequence & \multicolumn{2}{c}{\vthree{}-2D} & \multicolumn{2}{c}{MesonGS} & \multicolumn{2}{c}{LTS-Draco} & \multicolumn{2}{c}{G-PCC} & \multicolumn{2}{c}{\sysname{} (Ours)} \\
           & Enc & Dec & Enc & Dec & Enc & Dec & Enc & Dec & Enc & Dec \\
  \hline
  Actor1    & 443.1  & 312.5 & 144950.2 & 487.4 & 136.8  & 84.4  & 5063.9  & 2986.5  & 17.4 & 12.8 \\
  Actor2    & 4808.9 & 362.8 & 144363.5 & 368.8 & 103.5  & 60.8  & 3679.6  & 2205.4  & 15.6 & 12.0 \\
  Actor3    & 542.5  & 331.9 & 145057.5 & 529.3 & 156.3  & 95.1  & 6027.0  & 3474.0  & 17.6 & 13.2 \\
  Actor4    & 554.1  & 358.9 & 145351.6 & 622.6 & 182.8  & 104.6 & 8070.4  & 4858.2  & 19.1 & 15.0 \\
  Actor5    & 435.7  & 330.0 & 145003.4 & 559.3 & 158.2  & 91.9  & 5992.8  & 3693.7  & 17.8 & 13.4 \\
  Actor6    & 483.5  & 367.4 & 145344.3 & 674.9 & 201.9  & 113.1 & 8579.4  & 5163.9  & 19.6 & 15.7 \\
  Actor7    & 421.3  & 323.4 & 144602.4 & 504.7 & 151.8  & 88.2  & 5294.2  & 3174.0  & 17.7 & 13.2 \\
  \midrule
  Coffee Martini   & 473.6 & 352.0 & 30228.0 & 1415.6 & 1059.4 & 215.8 & 12540.2 & 10090.6 & 24.9 & 23.9 \\
  Cook Spinach     & 316.8 & 262.8 & 28719.5 & 1012.1 & 229.7  & 127.2 & 7467.1  & 5866.7  & 21.7 & 20.1 \\
  Cut Roasted Beef & 313.3 & 261.0 & 28781.4 & 978.2  & 229.4  & 128.5 & 8162.4  & 6438.3  & 21.5 & 19.6 \\
  Flame Salmon     & 488.0 & 340.3 & 29692.5 & 1335.9 & 413.8  & 214.0 & 11327.0 & 9236.8  & 27.7 & 23.4 \\
  Flame Steak      & 316.9 & 262.3 & 28877.8 & 1080.6 & 222.4  & 129.3 & 6364.7  & 4548.4  & 22.2 & 20.3 \\
  Sear Steak       & 312.8 & 263.2 & 28826.4 & 1061.6 & 247.9  & 133.3 & 6966.3  & 5391.3  & 22.2 & 20.5 \\
  \bottomrule
  \end{tabular}
\end{table*}

\section{\sysname{} Octree Encoding}\label{app:sysn-octr-encod}

Algorithm~\ref{algo:octree-encode-gpu} summarizes our GPU octree encoder. Starting from voxelized integer coordinates, we first compute Morton codes for all occupied leaf voxels, then radix-sort and deduplicate them to obtain the leaf set $L_J$. The upper octree levels are constructed bottom-up by repeatedly mapping child codes to their parents and removing duplicates, yielding level-wise node sets $\{L_d\}_{d=0}^J$.

After the hierarchy is built, we serialize a compact occupancy bitstream. The header stores the number of levels and the node count of each level. The payload is then generated in breadth-first order: for each parent in level $L_d$, we produce one occupancy byte indicating which of its eight children are present in $L_{d+1}$. As shown in Algorithm~\ref{algo:octree-encode-gpu}, this is done by locating the base child Morton code and setting the corresponding bits for all valid children in the next level.

This construction avoids explicit pointer-based octree data structures and operates directly on sorted Morton codes, which is well suited to parallel primitives such as radix sort, deduplication, and batched searches. The resulting bitstream is a level-synchronous octree representation that can be decoded with the same breadth-first traversal order.

\begin{algorithm}[t]
  \caption{GPU Octree{}'s Encoding from Voxelized Positions}
  \label{algo:octree-encode-gpu}
\SetAlgoLined
\DontPrintSemicolon

\SetKwInOut{Input}{Input}
\SetKwInOut{Output}{Output}

\SetKwFunction{Morton}{Morton}
\SetKwFunction{RadixSort}{RadixSort}
\SetKwFunction{Unique}{Unique}
\SetKwFunction{Parent}{Parent}
\SetKwFunction{OccKernel}{OccKernel}
\SetKwFunction{Concat}{Concat}
\SetKwFunction{Append}{Append}
\SetKwFunction{BinarySearch}{BinarySearch}

\Input{Voxelized integer positions $\hat{V}=\{(\hat{x}_i,\hat{y}_i,\hat{z}_i)\}_{i=1}^{\hat{N}_v}$ on GPU; octree depth $J$}
\Output{Serialized octree bitstream $B_{occ}$}

\BlankLine
\tcp*[l]{Step 1: Morton codes for leaf voxels}
\ForPar{$i \gets 1$ \KwTo $\hat{N}_v$}{
  $m_i \gets \Morton(\hat{x}_i,\hat{y}_i,\hat{z}_i, J)$
}
$L_J \gets \Unique(\RadixSort(\{m_i\}))$ \tcp*{leaf nodes}

\BlankLine
\tcp*[l]{Step 2: Level-synchronous octree construction}
\For(\tcp*[f]{bottom-up parents}){$d \gets J-1$ \KwTo $0$}{
  $L_d \gets \Unique(\{\, \Parent(c_m)\;|\;c_m\in L_{d+1}\,\})$ \tcp*{$\Parent(c_m)=c_m\gg 3$}
}

\BlankLine
\tcp*[l]{Step 3: Bitstream header}
$num\_levels \gets J+1$\;
$level\_sizes \gets (|L_0|,\ldots,|L_J|)$\;
$B_{occ} \gets \Append(B_{occ}, num\_levels, level\_sizes)$\;

\BlankLine
\tcp*[l]{Step 4: Breadth-first occupancy bitstream.}
\ForPar{$d \gets 0$ \KwTo $J-1$}{
  $B_d \gets \OccKernel(L_d, L_{d+1})$
  $B_{occ} \gets \Append(B_{occ}, B_d)$ \tcp*{contiguous bytes for level $d$}
}

\Return $B_{occ}$\;

\BlankLine
\textbf{Kernel \OccKernel($L_d,L_{d+1}$):}\;
\ForPar{$p \in L_d$}{
  $c_m^b \gets p \ll 3$ \tcp*{base child code}
  $occ \gets 0$ \tcp*{occupancy byte}
  $c_m \gets \BinarySearch(L_{d+1}, c_m^b)$
  \For{$k \gets 0$ \KwTo $7$}{
      \If{$L_{d+1}[c_m+k] \leq c_m^b+7$}{
        $occ \gets occ \;|\; \big(1 \ll (L_{d+1}[c_m+k]-c_m^b)\big)$\;
      }
      
  }
  Write $occ$ into $B_d$ at parent's index\;
}
\end{algorithm}

\section{\sysname{}'s Octree Decoding}\label{app:sysn-octr-decod}
\begin{algorithm}[t]
  \caption{GPU Octree Decoding}
  \label{algo:octree-decode-gpu}
\SetAlgoLined
\DontPrintSemicolon

\SetKwInOut{Input}{Input}
\SetKwInOut{Output}{Output}

\SetKwFunction{ReadHeader}{ReadHeader}
\SetKwFunction{ReadOcc}{ReadOccBytes}
\SetKwFunction{Popcount}{Popcount}
\SetKwFunction{ExScan}{ExclusiveScan}
\SetKwFunction{ExpandKernel}{ExpandChildrenKernel}
\SetKwFunction{ReconKernel}{ReconstructPointsKernel}
\SetKwFunction{MortonDec}{MortonDecode}
\SetKwFunction{Interleave}{InterleaveXYZ}

\Input{Compressed octree buffer $B$ on GPU; octree depth $J$}
\Output{Voxelized positions $\hat{V}$ on GPU; leaf Morton codes $L_J$ on GPU}
\BlankLine
\tcp*[l]{Step 1: Parse header.}
$(num\_levels,\ level\_sizes,\ ptr) \gets \ReadHeader(B)$\;

\BlankLine
\tcp*[l]{Step 2: Initialize root list (implicit root Morton code).}
$L_0 \gets [\,0\,]$

\BlankLine
\tcp*[l]{Step 3: Level-synchronous decoding.}
\For(\tcp*[f]{top-down levels}){$d \gets 0$ \KwTo $J-1$}{
  $M_d \gets |L_d|$\;
  $Occ_d \gets \ReadOcc(ptr, M_d)$ \tcp*{read $M_d$ bytes from $B$ into a GPU array}
  $ptr \gets ptr + M_d$\;

  \BlankLine
  \tcp*[l]{Count children per parent.}
  \ForPar{$p \gets 1$ \KwTo $M_d$}{
    $cnt[p] \gets \Popcount(Occ_d[p])$\;
  }

  \BlankLine
  \tcp*[l]{Find total children at $d+1$.}
  $off \gets \ExScan(cnt)$
  $T \gets off[M_d] + cnt[M_d]$\;
  Allocate $L_{d+1}$ of length $T$\;

  \BlankLine
  \tcp*[l]{Expand children.}
  \ExpandKernel$(L_d, Occ_d, off, L_{d+1})$ with $M_d$ threads\;
}

\BlankLine
\tcp*[l]{Step 6: Get voxel coordinates at leaf.}
$\hat{N}_v \gets |L_J|$\;
$shift \gets J_v - J$\;
Allocate $\hat{x}[1{:}\hat{N}_v],\hat{y}[1{:}\hat{N}_v],\hat{z}[1{:}\hat{N}_v]$\;
Launch \ReconKernel$(L_J, shift, \hat{x},\hat{y},\hat{z})$ with $\hat{N}_v$ threads\;
$\hat{V} \gets \Interleave(\hat{x},\hat{y},\hat{z})$ \tcp*{Reverse Morton Code.}

\Return $(\hat{V}, L_J)$\;
\end{algorithm}

\begin{algorithm}[h!]
  \caption{\texttt{expand\_children\_kernel}: Expand occupied children and generate Morton codes}
  \label{algo:octree-expand-children-kernel}
\SetAlgoLined
\DontPrintSemicolon
\SetKwInOut{Input}{Input}
\SetKwInOut{Output}{Output}

\Input{Parent Morton codes $L_d[1{:}M_d]$; occupancy bytes $Occ_d[1{:}M_d]$; write offsets $off[1{:}M_d]$}
\Output{Child Morton codes $L_{d+1}[1{:}T]$}

\ForPar{$p \gets 1$ \KwTo $M_d$}{
  $parent \gets L_d[p]$\;
  $occ \gets Occ_d[p]$\;
  $base \gets parent \ll 3$ \tcp*{base child Morton code}
  $w \gets off[p]$ \tcp*{exclusive-scan start offset for this parent}
  \For{$i \gets 0$ \KwTo $7$}{
    \If{$occ \wedge (1 \ll i)$}{
      $L_{d+1}[w] \gets base + i$\;
      $w \gets w + 1$\;
    }
  }
}
\end{algorithm}
The GPU decoder reconstructs voxelized geometry by replaying the serialized occupancy stream in a level-synchronous manner, without building any explicit tree pointers.
Starting from the implicit root Morton code ($L_0=\{0\}$), for each depth $d$ it reads the $|L_d|$ occupancy bytes, counts the number of occupied children per parent (population count), and performs an exclusive prefix-sum to compute disjoint write offsets for the next-level array.
It then launches \texttt{expand\_children\_kernel} (Alg.~\ref{algo:octree-expand-children-kernel}), which generates each occupied child Morton code in $O(1)$ via $(parent \ll 3)+i$ and writes all children contiguously into $L_{d+1}$, avoiding atomics despite variable fan-out.
Repeating this from $d=0$ to $J{-}1$ yields the leaf Morton codes $L_J$, which are finally decoded into integer voxel coordinates (and optionally interleaved) in a parallel post-pass.
Overall, decoding preserves the same breadth-first structure as the encoder while turning the key sequential dependency (unknown next-level size) into a standard GPU pattern: count $\rightarrow$ prefix-sum $\rightarrow$ parallel scatter.
Full algorithm can be found here~\ref{algo:octree-decode-gpu}.

\section{\sysname{}'s RAHT Implementation}\label{app:sysn-raht-encod}

We implement RAHT as a two-stage GPU pipeline: a \emph{prelude} that builds the merge schedule from Morton-sorted occupied voxels, and a \emph{transform} stage that applies the weighted orthonormal rotations using that schedule. This separation isolates the octree-dependent control flow from the per-attribute transform, allowing the latter to be executed with simple parallel kernels.

Algorithm~\ref{algo:raht-prelude} gives the reference non-parallel prelude. For each of the $3J$ binary merge steps induced by a depth-$J$ octree, it constructs the active index list $I_\ell$, the corresponding run-length weights $W_\ell$, and the sibling flags $F_\ell$. Here, $W_\ell$ records the number of original voxels represented by each active entry, and $F_\ell$ marks whether an entry is the left element of a valid sibling pair. The sibling relation is tested directly from Morton codes using XOR and a level-dependent mask, without explicitly materializing the octree.

Algorithm~\ref{algo:raht-prelude-gpu} shows the GPU version of this schedule construction. At each level, one thread computes the weight and sibling flag for one active entry. The next active list is then obtained by removing right siblings while keeping left siblings and singletons. This produces the schedule $[(I_\ell, W_\ell, F_\ell)]_{\ell=1}^{3J}$ used by the transform.

Algorithm~\ref{algo:raht-forward} applies the forward or inverse RAHT on the GPU. In the forward pass, levels are processed bottom-up; in the inverse pass, they are processed in reverse order. For each sibling pair, the transform uses the standard RAHT weights
$$
a=\sqrt{\frac{w_0}{w_0+w_1}}, \qquad
b=\sqrt{\frac{w_1}{w_0+w_1}},
$$
to perform the corresponding orthonormal rotation independently for each attribute channel. Since pairs within a level are disjoint, the computation can be parallelized across active entries.

\begin{algorithm}[h!]
  \caption{RAHT prelude (non-parallel)}
  \label{algo:raht-prelude}
\SetAlgoLined
\DontPrintSemicolon
\SetKwInOut{Input}{Input}
\SetKwInOut{Output}{Output}

\Input{Morton codes $M[0{:}N_v{-}1]$, Num voxels $N_v$, octree depth $J$}
\Output{$[(I_\ell, W_\ell, F_\ell)]_{\ell=1}^{3J}$}

\For{$\ell \gets 1$ \KwTo $3J$}{
  \uIf{$\ell = 1$}{
    $I_1 \gets (0{:}N_v{-}1)^{T}$ \tcp*{active indices at level 1}
  }\Else{
    $I_\ell \gets I_{\ell-1}\big(\neg [0; F_{\ell-1}]\big)$ \tcp*{keep left siblings + singletons}
  }
  $M_\ell \gets M[I_\ell]$ \tcp*{Morton codes at level $\ell$}
  $W_\ell \gets [I_\ell(2{:}\mathrm{end});\, N_v] - I_\ell$ \tcp*{run-length weights (sentinel $N_v$)}
  $D \gets M_\ell(1{:}\mathrm{end}{-}1)\ \oplus\ M_\ell(2{:}\mathrm{end})$ \tcp*{path diffs}
  $F_\ell \gets \big(D\ \wedge\ (2^{3J} - 2^{\ell})\big) = 0$ \tcp*{left-sibling flags}
}
\Return $[(I_\ell, W_\ell, F_\ell)]_{\ell=1}^{3J}$\;
\end{algorithm}

\begin{algorithm}[t]
  \caption{RAHT Prelude on GPU}
  \label{algo:raht-prelude-gpu}
\SetAlgoLined
\DontPrintSemicolon
\SetKwInOut{Input}{Input}
\SetKwInOut{Output}{Output}
\SetKwFunction{Shift}{torch.shift}
\SetKwFunction{MaskedSelect}{torch.masked\_select}
\SetKwFunction{FWKernel}{FWKernel}

\Input{Morton codes $M[0{:}N_v{-}1]$, Num voxels $N_v$, octree depth $J$}
\Output{$[(I_\ell, W_\ell, F_\ell)]_{\ell=1}^{3J}$}

$I_1 \gets (0{:}N_v{-}1)^{T}$\;
\For{$\ell \gets 1$ \KwTo $3J$}{
  $\mu_\ell \gets (2^{3J} - 2^{\ell})$ \tcp*{mask for sibling test}
  \ForPar(\tcp*[f]{1 GPU thread per $k$}){$k \gets 1$ \KwTo $M_\ell$}{
    $curr \gets I_\ell(k)$\;
    $next \gets \begin{cases}
      end & \text{if } k = M_\ell\\
      I_\ell(k{+}1) & \text{otherwise}
    \end{cases}$\;

    $W_\ell(k) \gets next - curr$\;

    \uIf{$k = M_\ell$}{
      $F_\ell(k) \gets 0$\;
    }\Else{
      $F_\ell(k) \gets ((M(curr) \oplus M(next)) \wedge \mu_\ell ) = 0$\;
    }
  }

  $P_\ell \gets \Shift(F_\ell)$ \tcp*{$[0; F_\ell(0{:}\mathrm{end}{-}1)]$ marks right siblings}
  $I_{\ell+1} \gets \MaskedSelect(I_\ell, \neg P_\ell)$

  \If{$|I_{\ell+1}| = 1$}{
    \textbf{break}\;
  }
}
\Return $[(I_\ell, W_\ell, F_\ell)]_{\ell=1}^{3J}$\;
\end{algorithm}






\begin{algorithm}[h!]
\caption{RAHT Transform (Forward / Inverse) on GPU}
\label{algo:raht-forward}
\SetAlgoLined
\DontPrintSemicolon
\SetKwInOut{Input}{Input}
\SetKwInOut{Output}{Output}

\Input{Attributes $A\in\mathbb{R}^{N_v\times C}$ (GPU), prelude schedule $[(I_\ell,W_\ell,F_\ell)]_{\ell=1}^{3J}$, \texttt{inverse} flag}
\Output{Coefficients $Y\in\mathbb{R}^{N_v\times C}$ (GPU)}

$Y \gets A$ \tcp*{clone for in-place updates}
\If{\texttt{inverse} = 0}{
  \For(\tcp*[f]{bottom-up}){$\ell \gets 1$ \KwTo $3J-1$}{
    $M_\ell \gets |I_\ell|$\;
    Launch \texttt{raht\_forward\_level\_kernel} with $M_\ell$ threads\;
    \Indp
    \ForPar{$k \gets 0$ \KwTo $M_\ell-2$}{
      \If{$F_\ell[k]$}{
        $i_0 \gets I_\ell[k]$;\;
        $i_1 \gets I_\ell[k+1]$;\;
        $w_0 \gets W_\ell[k]$;\;
        $w_1 \gets W_\ell[k+1]$;\;
        $a \gets \sqrt{w_0/(w_0+w_1)}$;\;
        $b \gets \sqrt{w_1/(w_0+w_1)}$;\;
        \For{$c \gets 0$ \KwTo $C-1$}{
          $y_0 \gets Y[i_0,c]$;\;
          $y_1 \gets Y[i_1,c]$;\;
          $Y[i_0,c] \gets a\cdot y_0 + b\cdot y_1$\;
          $Y[i_1,c] \gets -b\cdot y_0 + a\cdot y_1$\;
        }
      }
    }
    \Indm
  }
}{
  \For(\tcp*[f]{top-down}){$\ell \gets 3J-1$ \KwTo $1$}{
    $M_\ell \gets |I_\ell|$\;
    Launch \texttt{raht\_inverse\_level\_kernel} with $M_\ell$ threads\;
    \Indp
    \ForPar{$k \gets 0$ \KwTo $M_\ell-2$}{
      \If{$F_\ell[k]$}{
        $i_0 \gets I_\ell[k]$;\;
        $i_1 \gets I_\ell[k+1]$;\;
        $w_0 \gets W_\ell[k]$;\;
        $w_1 \gets W_\ell[k+1]$;\;
        $a \gets \sqrt{w_0/(w_0+w_1)}$;\;
        $b \gets \sqrt{w_1/(w_0+w_1)}$;\;
        \For{$c \gets 0$ \KwTo $C-1$}{
          $y_0 \gets Y[i_0,c]$;\;
          $y_1 \gets Y[i_1,c]$;\;
          $Y[i_0,c] \gets a\cdot y_0 - b\cdot y_1$\;
          $Y[i_1,c] \gets b\cdot y_0 + a\cdot y_1$\;
        }
      }
    }
    \Indm
  }
}
\Return $Y$\;
\end{algorithm}

\section{R-D Curves}

\begin{figure}[htbp]
    \centering
    \begin{subfigure}[b]{0.48\linewidth}
        \centering
        \includegraphics[width=\linewidth]{figures/rd_baselines/rd_baselines_curve_4K_Actor1_Greeting_frame0.pdf}
        \caption{Actor1\_Greeting}
        \label{fig:rd-actor1-greeting}
    \end{subfigure}
    \hfill
    \begin{subfigure}[b]{0.48\linewidth}
        \centering
        \includegraphics[width=\linewidth]{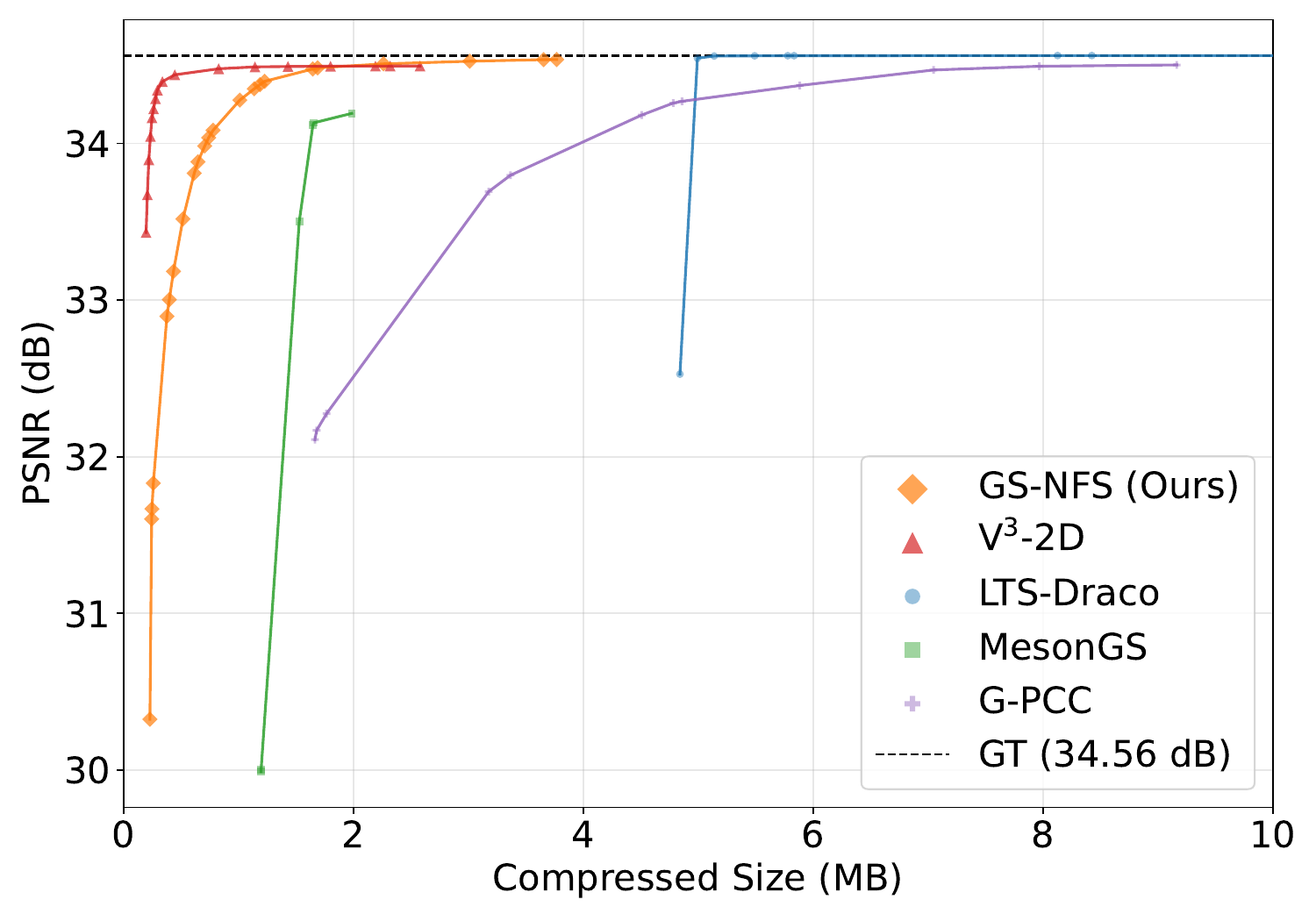}
        \caption{Actor2\_Dancing}
        \label{fig:rd-actor2-dancing}
    \end{subfigure}
    \vspace{0.5em}
    \begin{subfigure}[b]{0.48\linewidth}
        \centering
        \includegraphics[width=\linewidth]{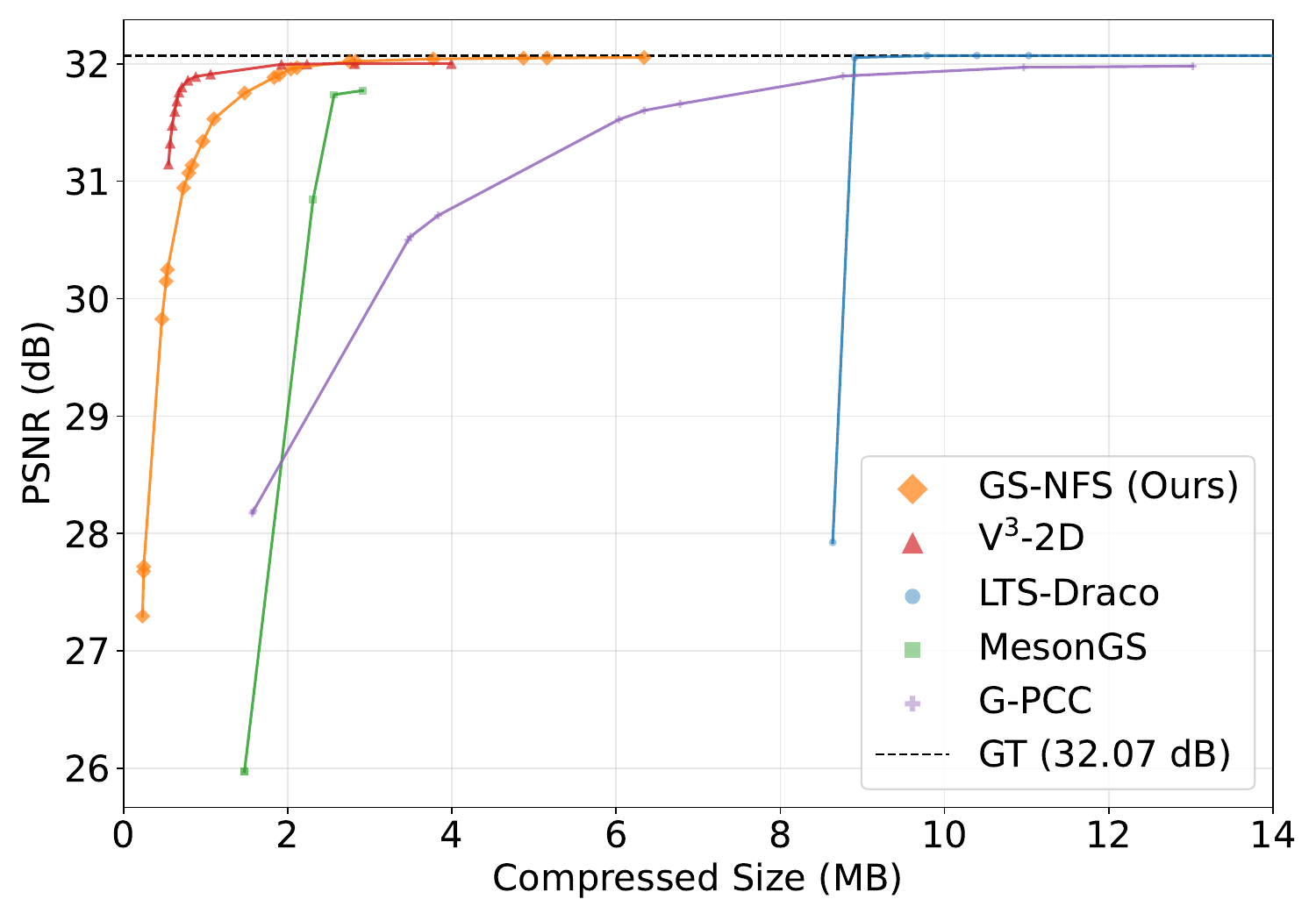}
        \caption{Actor3\_Violin}
        \label{fig:rd-actor3-violin}
    \end{subfigure}
    \hfill
    \begin{subfigure}[b]{0.48\linewidth}
        \centering
        \includegraphics[width=\linewidth]{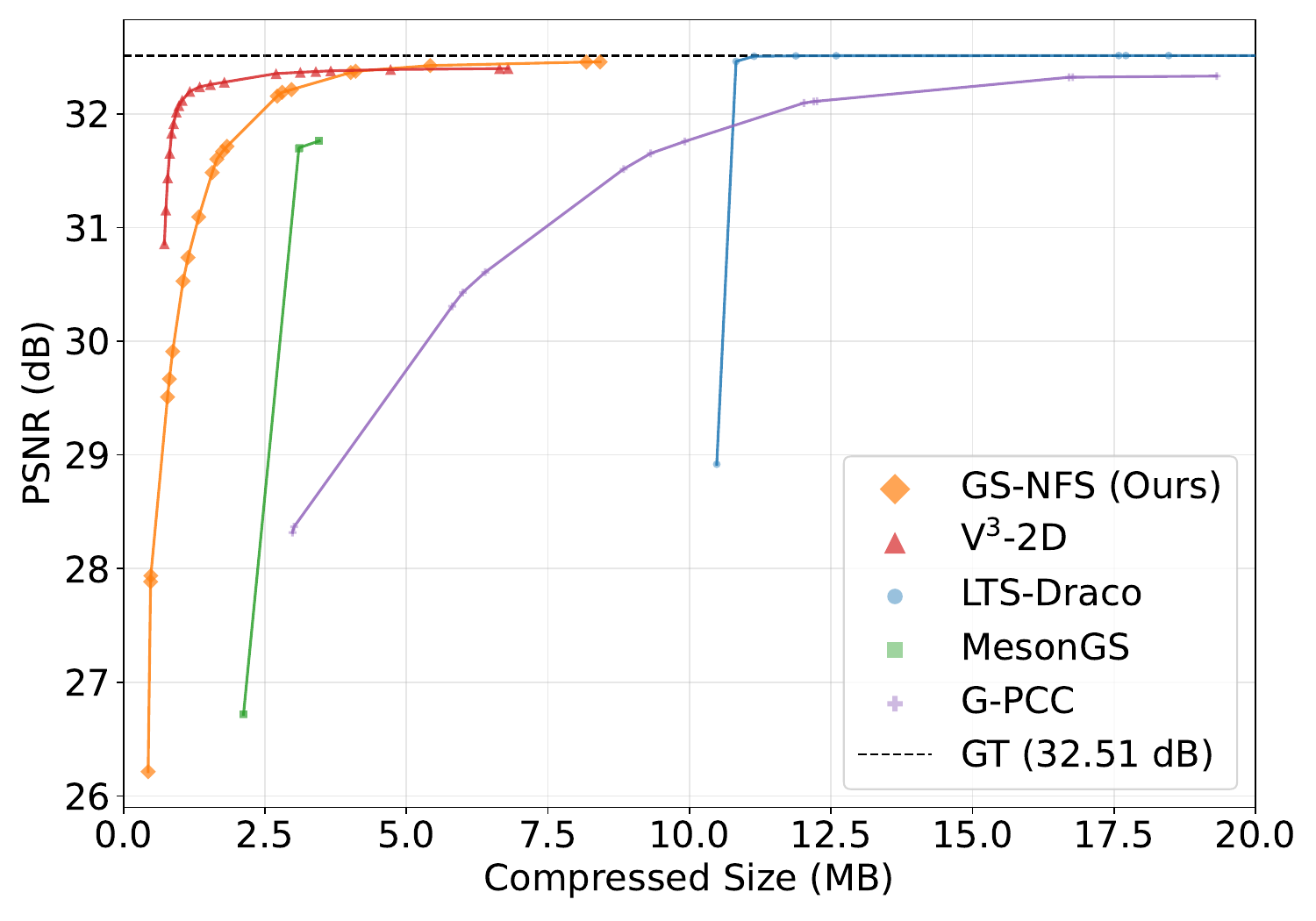}
        \caption{Actor4\_Dancing}
        \label{fig:rd-actor4-dancing}
    \end{subfigure}
    \vspace{0.5em}
    \begin{subfigure}[b]{0.48\linewidth}
        \centering
        \includegraphics[width=\linewidth]{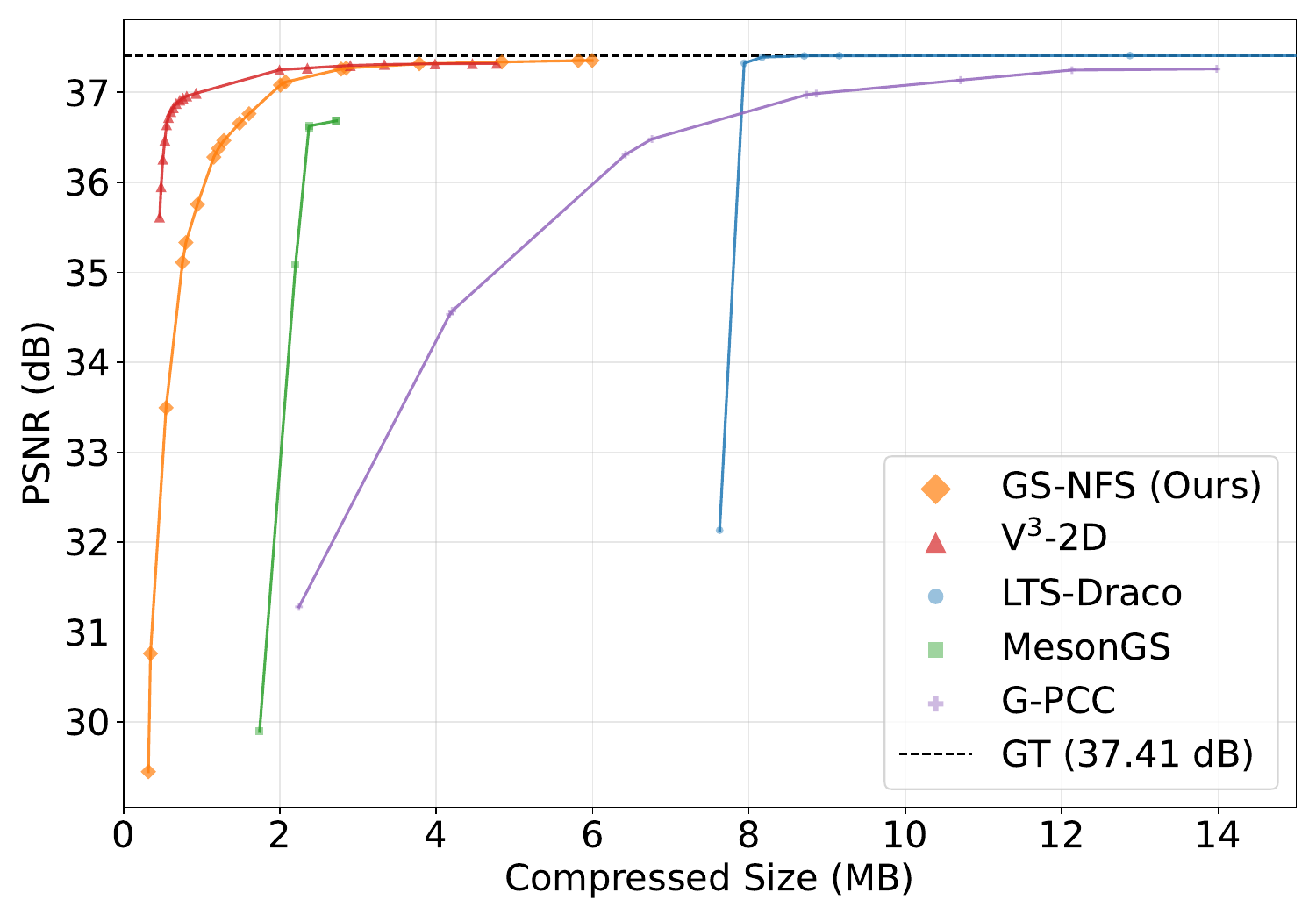}
        \caption{Actor5\_Oil-paper\_Umbrella}
        \label{fig:rd-actor5-umbrella}
    \end{subfigure}
    \hfill
    \begin{subfigure}[b]{0.48\linewidth}
        \centering
        \includegraphics[width=\linewidth]{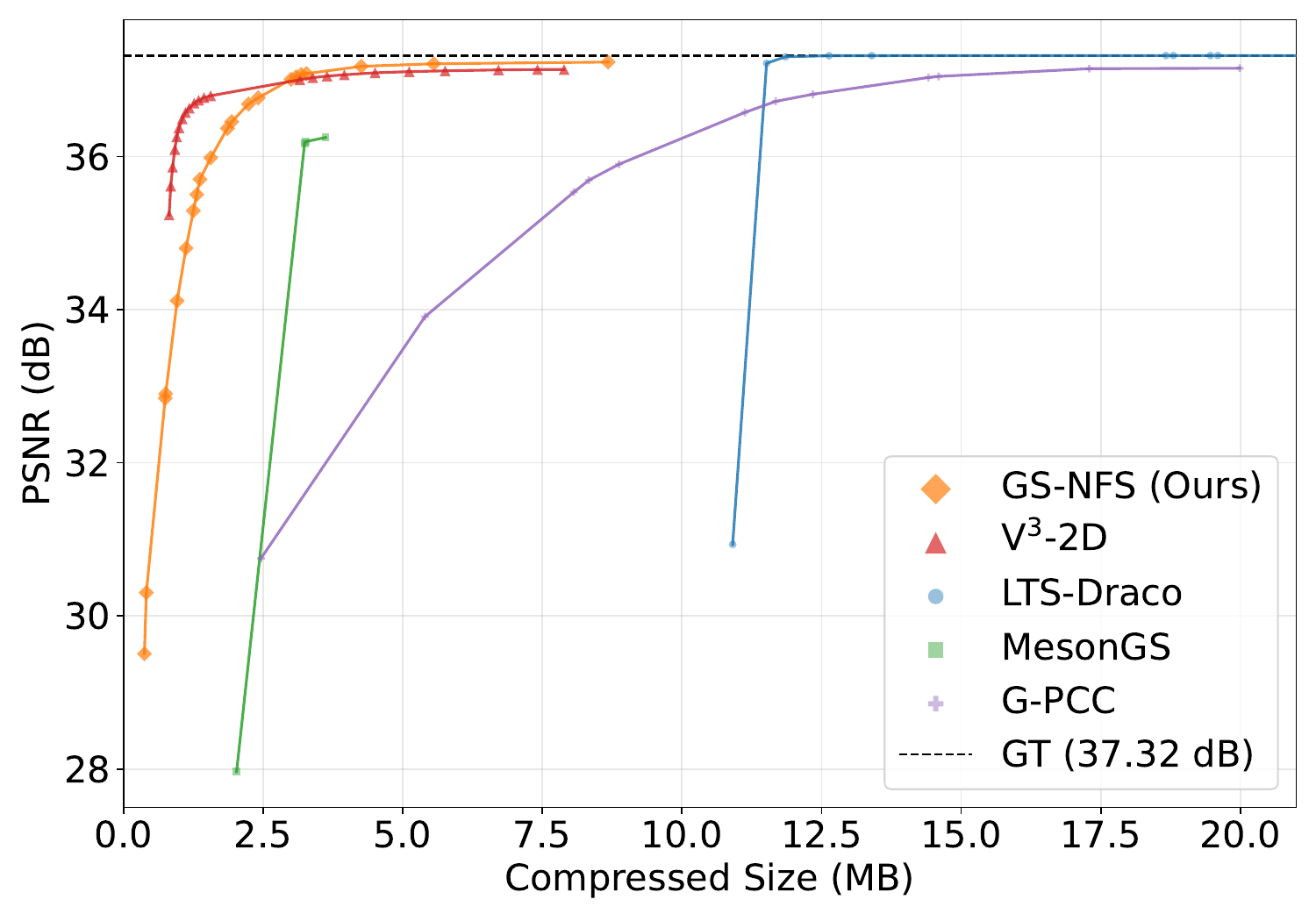}
        \caption{Actor6\_Changing\_Clothes}
        \label{fig:rd-actor6-clothes}
    \end{subfigure}
    \vspace{0.5em}
    \begin{subfigure}[b]{0.48\linewidth}
        \centering
        \includegraphics[width=\linewidth]{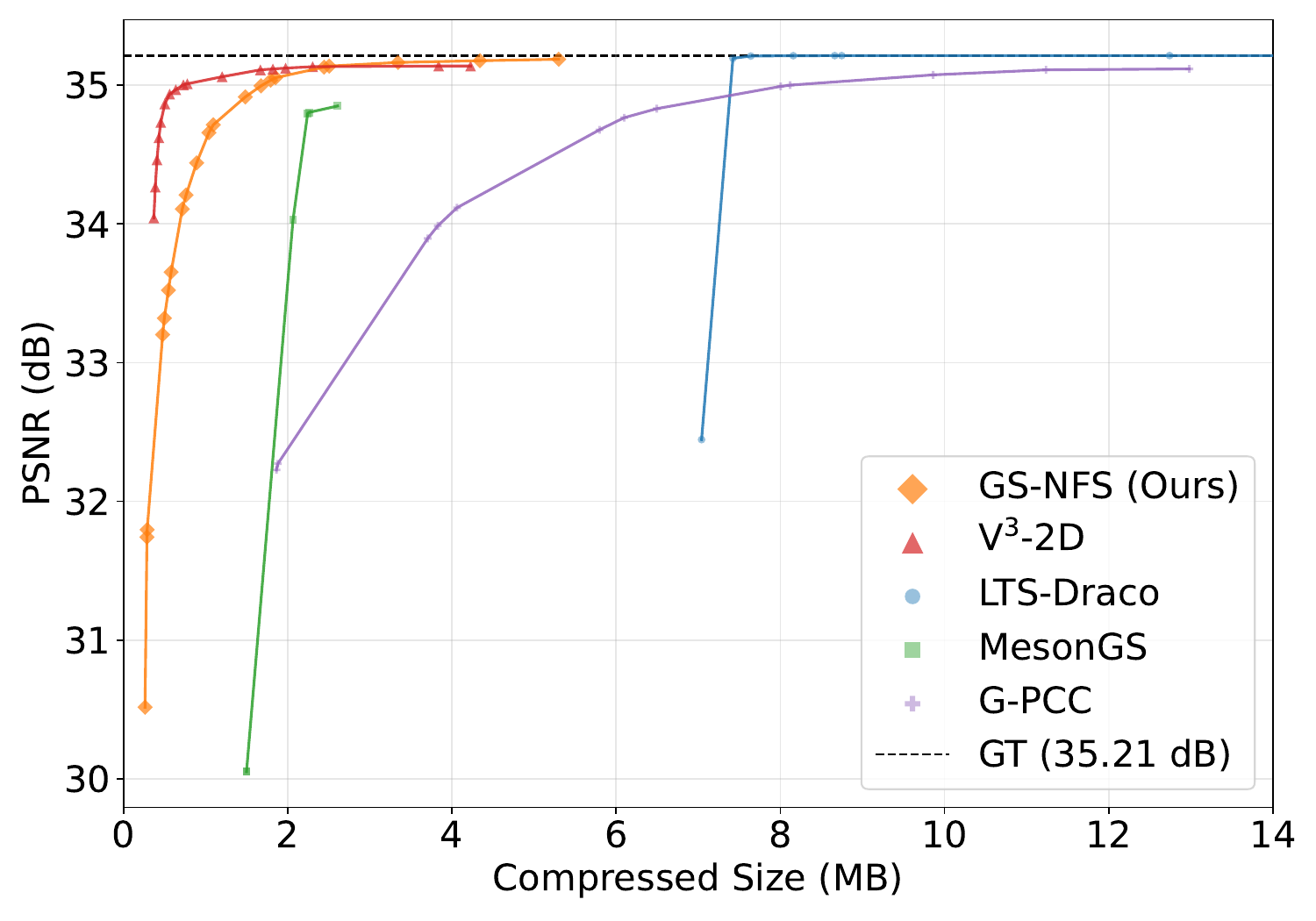}
        \caption{Actor7\_Nunchaku}
        \label{fig:rd-actor7-nunchaku}
    \end{subfigure}
    \caption{Rate-distortion curves for HiFi4G 4K sequences (frame 0).}
    \label{fig:rd-baselines-hifi4g}
\end{figure}

\begin{figure}[htbp]
    \centering
    \begin{subfigure}[b]{0.48\linewidth}
        \centering
        \includegraphics[width=\linewidth]{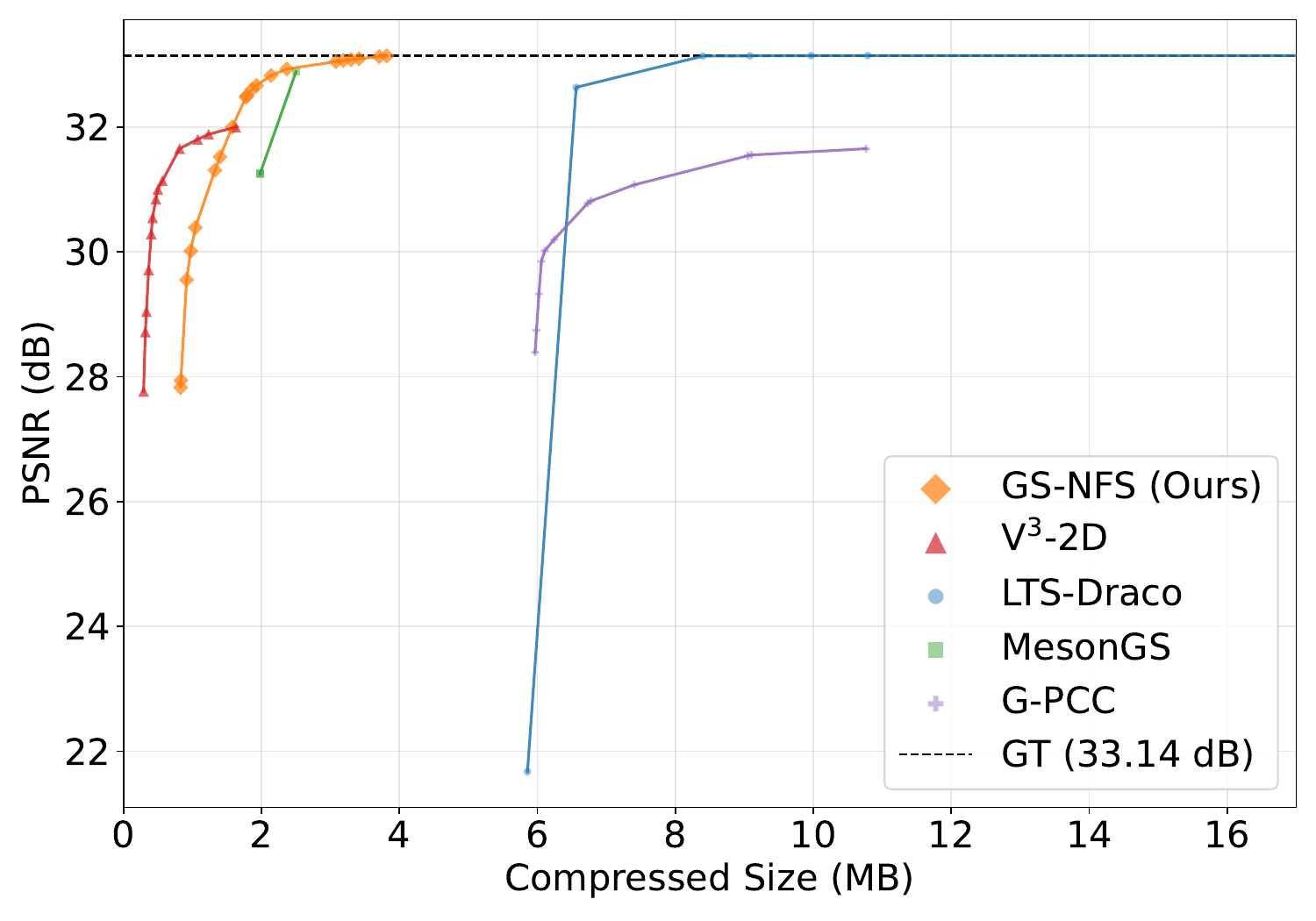}
        \caption{cook\_spinach}
        \label{fig:rd-cook-spinach}
    \end{subfigure}
    \hfill
    \begin{subfigure}[b]{0.48\linewidth}
        \centering
        \includegraphics[width=\linewidth]{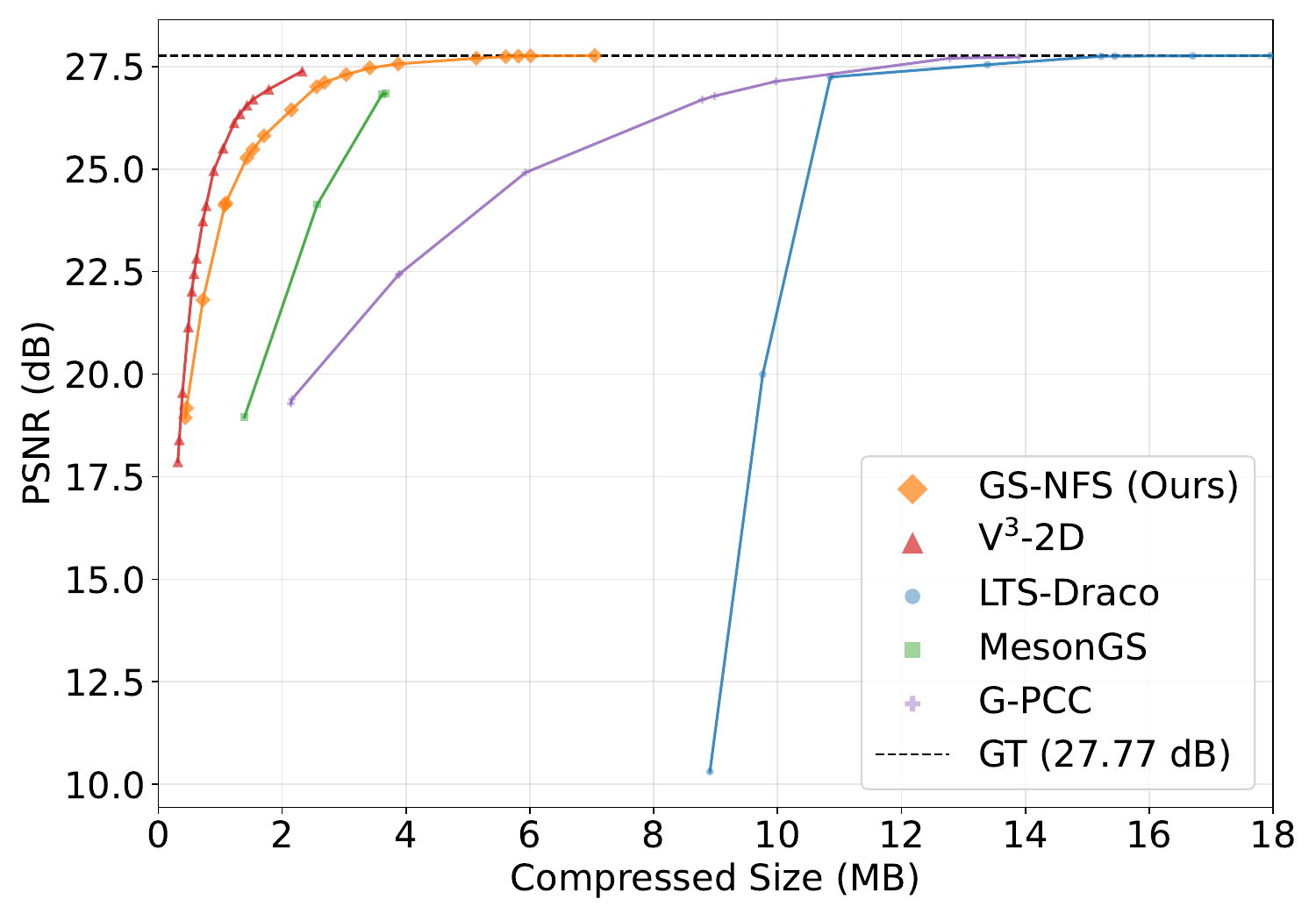}
        \caption{coffee\_martini}
        \label{fig:rd-coffee-martini}
    \end{subfigure}
    \vspace{0.5em}
    \begin{subfigure}[b]{0.48\linewidth}
        \centering
        \includegraphics[width=\linewidth]{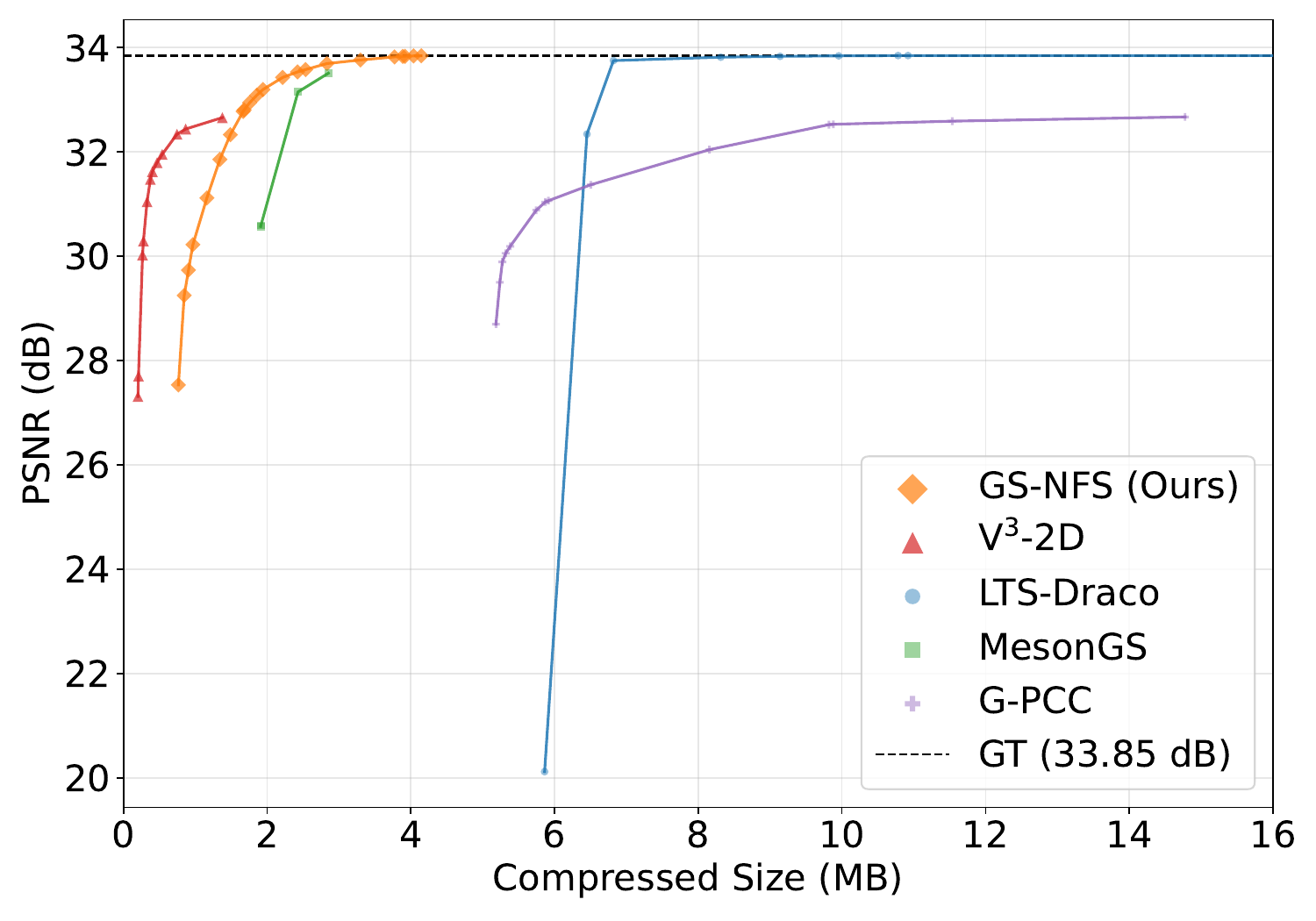}
        \caption{cut\_roasted\_beef}
        \label{fig:rd-cut-roasted-beef}
    \end{subfigure}
    \hfill
    \begin{subfigure}[b]{0.48\linewidth}
        \centering
        \includegraphics[width=\linewidth]{figures/rd_baselines/rd_baselines_curve_flame_salmon_1_frame1.pdf}
        \caption{flame\_salmon\_1}
        \label{fig:rd-flame-salmon}
    \end{subfigure}
    \vspace{0.5em}
    \begin{subfigure}[b]{0.48\linewidth}
        \centering
        \includegraphics[width=\linewidth]{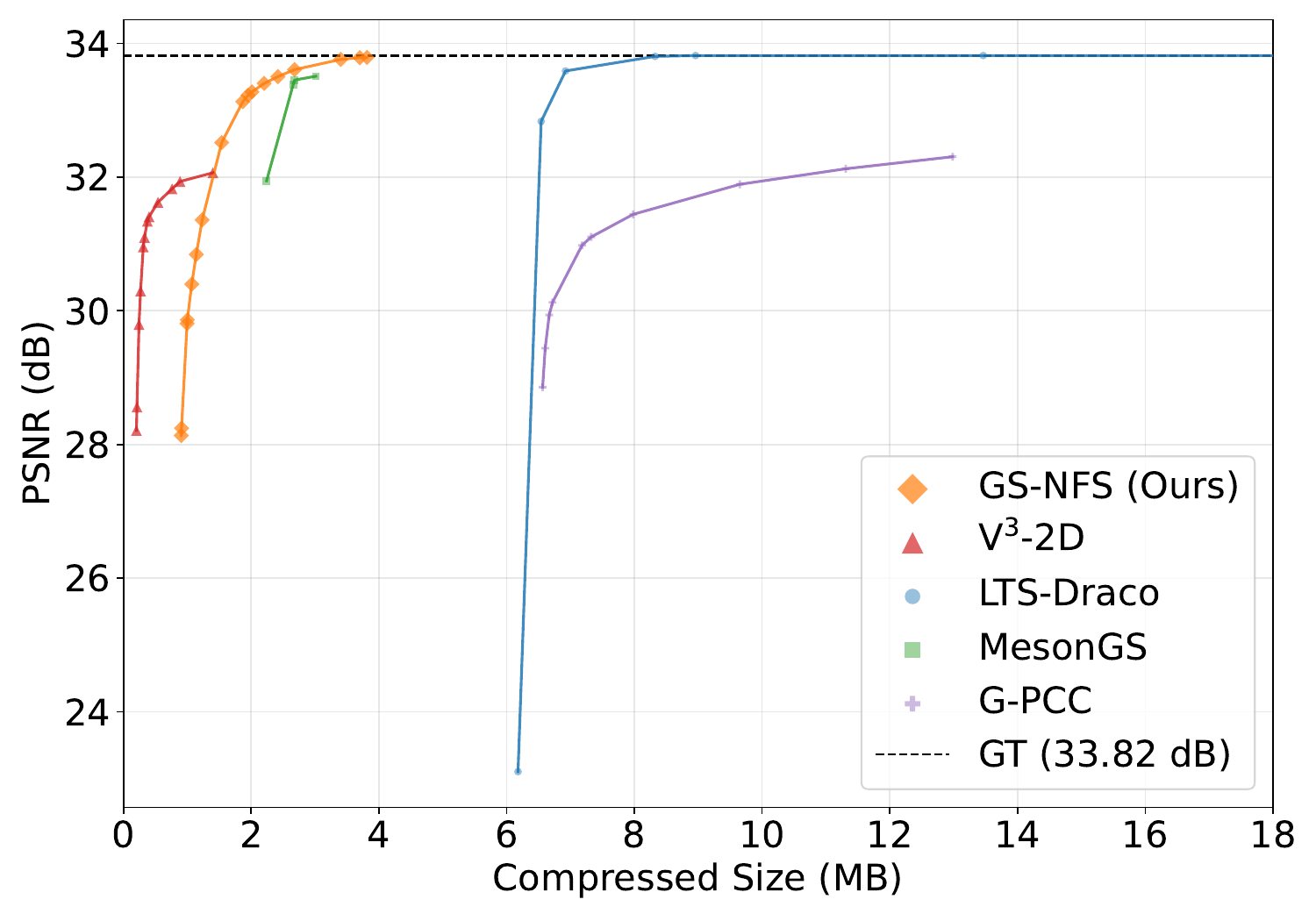}
        \caption{flame\_steak}
        \label{fig:rd-flame-steak}
    \end{subfigure}
    \hfill
    \begin{subfigure}[b]{0.48\linewidth}
        \centering
        \includegraphics[width=\linewidth]{figures/rd_baselines/rd_baselines_curve_sear_steak_frame1.pdf}
        \caption{sear\_steak}
        \label{fig:rd-sear-steak}
    \end{subfigure}
    \caption{Rate-distortion curves for N3DV sequences (frame 1).}
    \label{fig:rd-baselines-n3dv}
\end{figure}

This section provides the per-sequence R-D curves for all evaluated methods. For completeness, we include one representative frame from each sequence and plot the corresponding bitrate--quality trade-off. Figures~\ref{fig:rd-baselines-hifi4g} and~\ref{fig:rd-baselines-n3dv} report the results for the HiFi4G and N3DV datasets, respectively.


\section{Generating R-D Curves}\label{app:generating-r-d}


To generate the rate-distortion (R-D) curves, we perform an exhaustive parameter sweep for each baseline codec, per frame.

For each parameter configuration, we compress and decompress the 3DGS representation of a single frame, render the decompressed Gaussians, and compute PSNR against the ground-truth renders (rendered from the uncompressed model).

All renders use a resolution-downscale factor of ~2.

For \vthree{}-2D, which operates on groups of frames, we compress a group of 20 consecutive frames starting from the target frame and report the per-frame average compressed size; PSNR and SSIM are measured only on the target frame. 
All other baselines compress and evaluate a single frame independently.

The resulting (compressed size, quality) operating points are reduced to their upper convex hull to produce the final R-D curve.

Table~\ref{tab:rd-params} lists the swept parameters and their ranges for each baseline. Dataset-specific settings are noted where applicable.

\begin{table}[t]
\centering
\setlength{\tabcolsep}{3pt}
\small
\begin{tabular}{@{}llp{2.8cm}r@{}}
\toprule
\textbf{Method} & \textbf{Parameter} & \textbf{Values} & \textbf{\#} \\
\midrule
\multirow{2}{*}{V$^3$\!-2D}
  & QP         & $\{0, 1, \ldots, 40\}$           & \multirow{2}{*}{41} \\
  & Group size & 20                               & \\
\midrule
\multirow{4}{*}{MesonGS}
  & Octree depth & $\{8,10,12\}$/$\{12,14,16\}^\ast$ & \multirow{4}{*}{24} \\
  & Num.\ bits   & $\{8, 16\}$                       & \\
  & Blocks       & $\{57, 66\}$                      & \\
  & Codebook     & $\{2048, 4096\}$                  & \\
\midrule
\multirow{5}{*}{\shortstack[l]{LTS-\\Draco}}
  & \texttt{eg} (geom.)    & $\{0,8,12,16\}$        & \multirow{5}{*}{256} \\
  & \texttt{eo} (opacity)  & $\{0,8,12,16\}$        & \\
  & \texttt{et} (SH)       & $\{0,8,12,16\}$        & \\
  & \texttt{es} (scale)    & $\{0,8,12,16\}$        & \\
  & Comp.\ level           & 10                     & \\
\midrule
\multirow{2}{*}{G-PCC}
  & Octree depth & $\{8\text{--}12\}$/$\{12\text{--}17\}^\ast$ & \multirow{2}{*}{\shortstack[r]{225/\\270$^\ast$}} \\
  & $(q_r,q_d,q_o)^\dagger$ & 45 selected triples           & \\
\bottomrule
\end{tabular}
\vspace{2pt}
\raggedright
{\scriptsize $^\ast$\,HiFi4G\,/\,N3DV. G-PCC total = octree depths $\times$ 45 QP triples.\\
$^\dagger$\,$q_r$: QP for remaining attributes, $q_d$: QP for DC coefficients, $q_o$: QP for opacity.}
\caption{Parameter sweep configurations for R-D curve generation.}
\label{tab:rd-params}
\end{table}

\section{Compressing Static Scenes}\label{app:compr-stat-scen}

\begin{table}[t]
  \centering
  \footnotesize
  \scalebox{0.9}{
  \begin{tabular}{l|l|rr|rrr}
    \hline
    \textbf{Scene} & \textbf{Method} & \textbf{Enc.} & \textbf{Dec.} & \textbf{PSNR} & \textbf{Size} \\
    & & (ms) & (ms) & (dB) & (MB) \\
    \hline\hline
    \multirow{4}{*}{\textit{drjohnson}}
      & LTS-Draco            & 5,084    & 2,610   & 35.5 & 312.8 \\
      & MesonGS              & 220,066  & 22,421  & 32.6 & 40.6 \\
      & \textbf{\sysname{}}  & \textbf{165} & \textbf{183} & 34.4 & 37.8 \\
    \hline
    \multirow{4}{*}{\textit{truck}}
      & LTS-Draco            & 3,385    & 1,792   & 23.1 & 204.0 \\
      & MesonGS              & 319,395  & 14,357  & 22.7 & 27.8 \\
      & \textbf{\sysname{}}  & \textbf{91} & \textbf{108} & 22.4 & 19.0 \\
    \hline
  \end{tabular}
  }
  \caption{Static \gs{} scene compression. PSNR, SSIM, and LPIPS computed on rendered test views.}
  \label{tab:static-scenes}
\end{table}

\parab{Static \gs{} scene compression.}
We evaluate \sysname{} on two large static outdoor scenes from the Deep Blending dataset (\textit{drjohnson}, ~3M Gaussians, 748~MB) and Tanks \& Temples (\textit{truck}, ~2M Gaussians, 485~MB).
%
%
\cref{tab:static-scenes} shows the results.
\sysname{} achieves encode times of $91$--$165$~ms, which is \textbf{$\boldsymbol{31\times}$ faster} than LTS-Draco and \textbf{$\boldsymbol{1{,}200}$--$\boldsymbol{3{,}400\times}$ faster} than MesonGS.
At quality comparable to LTS-Draco ($<$1~dB PSNR difference), \sysname{} produces compressed files that are $5$--$10\times$ smaller.
%

\section{Other Ablations}\label{app:other-ablations}

\begin{table}[t]
  \centering
  \footnotesize
  \scalebox{0.9}{
  \begin{tabular}{l|l|rr|r}
    \hline
    \textbf{Sequence} & \textbf{Method} & \textbf{Enc. (ms)} & \textbf{Dec. (ms)} & \textbf{Size (KB)} \\
    \hline\hline
    \multirow{3}{*}{longdress}
      & Draco (CPU)       & 126  & 57  & 395 \\
      & G-PCC (CPU)        & 195  & 97  & 195 \\
      & \sysname{} octree & \textbf{2.9}  & \textbf{1.9} & 238 \\
    \hline
    \multirow{3}{*}{soldier}
      & Draco (CPU)       & 170  & 79  & 533 \\
      & G-PCC (CPU)        & 268  & 136 & 273 \\
      & \sysname{} octree & \textbf{3.4}  & \textbf{2.4} & 332 \\
    \hline
    \multirow{3}{*}{band2}
      & Draco (CPU)       & 123  & 54  & 561 \\
      & G-PCC (CPU)        & 325  & 213 & 526 \\
      & \sysname{} octree & \textbf{3.3}  & \textbf{2.6} & 565 \\
    \hline
    \multirow{3}{*}{pizza1}
      & Draco (CPU)       & 137  & 59  & 617 \\
      & G-PCC (CPU)        & 343  & 222 & 572 \\
      & \sysname{} octree & \textbf{3.3}  & \textbf{2.6} & 616 \\
    \hline
  \end{tabular}
  }
  \caption{Position-only octree encoding on point-cloud sequences. Encode/decode in ms, size in KB, averaged per frame.}
  \label{tab:ptcl-octree}
\end{table}

\parab{GPU vs.\ CPU octree encoding.}
We compare \sysname{}'s GPU octree encoder against Draco~\cite{draco} and G-PCC~\cite{gpcc} for position-only compression.
%
%
\cref{tab:ablation-octree} shows that \sysname{}'s GPU octree encoding time is \textbf{16-34$\boldsymbol\times$ faster} ($8$--$14\times$ for decoding) than Draco and \textbf{70--100$\boldsymbol\times$ faster} than G-PCC for encoding ($50$--$65\times$ for decoding), while producing bitstreams of comparable size.
ANS entropy coding of the octree occupancy bytes adds only ${\sim}0.3$~ms but roughly halves the compressed size.

\parab{RLGR parallelization.}
\begin{table}[t]
  \centering
  \footnotesize
  \begin{tabular}{l|rr|r|r}
    \hline
    \textbf{Variant} & \textbf{Block size} & \textbf{Enc.\ (ms)} & \textbf{Dec.\ (ms)} & \textbf{$\Delta$Size} \\
    \hline\hline
    CPU &  --   & 143.8  & 207.1 & 0.0\% \\
    GPU &  --   & 494.2  & 447.4 & 0.0\% \\
    GPU & 8192  & 15.7  & 15.0   & 0.1\% \\
    GPU & 4096  & 8.0   & 7.6    & 0.2\% \\
    GPU & 2048  & 4.1   & 4.0    & 0.4\% \\
    GPU & 1024  & 2.9   & 2.4    & 0.8\% \\
    GPU & 512   & 2.5   & 1.9    & 1.5\% \\
    GPU & 256   & 2.5   & 2.1    & 2.5\% \\
    \hline
  \end{tabular}
  \caption{RLGR entropy coding latency: CPU vs.\ GPU with varying block sizes for flame\_salmon.}
  \label{tab:ablation-rlgr}
\end{table}
\sysname{} parallelizes RLGR entropy coding by partitioning each attribute channel into fixed-size blocks and encoding them independently on GPU threads (\cref{sec:quant-entr-coding}).
\cref{tab:ablation-rlgr} compares the CPU baseline (sequential per-channel RLGR) against GPU variants with different block sizes.
Parallel RLGR a block size of $512$ achieves a \textbf{$\boldsymbol{58\times}$ encode} and \textbf{$\boldsymbol{109\times}$ decode speedup} on a desktop GPU over CPU, with only ${\sim}1.5$\% increase in compressed size.
On Jetson Orin, a block size of $512$ achieves ${\sim}6$~ms decode time per frame, compared to ${\sim}180$~ms for the CPU implementation--a $30\times$ speedup.
Smaller block sizes below 512 starts having diminishing returns.
Block size 2048--8192 provides a balanced tradeoff: $18$--$51\times$ speedup in encode/decode latency with negligible size overhead.

\begin{table}[t]
  \centering
  \footnotesize
  \scalebox{0.9}{
  \begin{tabular}{l|l|rr|r}
    \hline
    \textbf{Sequence} & \textbf{Method} & \textbf{Enc.\ (ms)} & \textbf{Dec.\ (ms)} & \textbf{Comp Ratio} \\
    \hline\hline
    \multirow{4}{*}{\begin{tabular}[c]{@{}l@{}}flame\_salmon\\ (400K Gauss)\end{tabular}}
      & Draco (CPU)                 & 98.0          & 37.5            & 10.9$\times$ \\
      & G-PCC (CPU)                  & 228.6         & 127.5           & 11.3$\times$ \\
      & Octree (w/o ANS) & 2.9           & 2.4           & 6.6$\times$ \\
      & Octree (w/ ANS)  & \textbf{3.2}  & \textbf{2.7}  & 11.2$\times$ \\
    \hline
    \multirow{4}{*}{\begin{tabular}[c]{@{}l@{}}sear\_steak\\ (250K Gauss)\end{tabular}}
      & Draco (CPU)                 & 64.9          & 28.1          & 5.1$\times$ \\
      & G-PCC (CPU)                  & 294.8         & 204.3         & 5.2$\times$ \\
      & Octree (w/o ANS) & 2.9           & 3.0           & 2.5$\times$ \\
      & Octree (w/ ANS)  & \textbf{3.2}  & \textbf{3.1}  & 4.8$\times$ \\
    \hline
    \multirow{4}{*}{\begin{tabular}[c]{@{}l@{}}Actor1\\ (130K Gauss)\end{tabular}}
      & Draco (CPU)                     & 30.7          & 13.0          & 6.7$\times$ \\
      & G-PCC (CPU)                      & 127.8         & 91.2          & 6.8$\times$ \\
      & Octree (w/o ANS)     & 1.8           & 1.4           & 3.4$\times$ \\
      & Octree (w/ ANS)      & \textbf{1.9}  & \textbf{1.6}  & 6.1$\times$ \\
    \hline
  \end{tabular}}
  \caption{Octree compression: CPU-based methods vs \sysname{}'s octree codec.}
  \label{tab:ablation-octree}
\end{table}

\begin{table}[t]
  \centering
  \footnotesize
  \scalebox{0.8}{
  \begin{tabular}{l|l|r|rr|r}
    \hline
    \textbf{Sequence} & \textbf{SH degree} & \textbf{Tot. ch} & \textbf{PyTorch (ms)} & \textbf{CUDA (ms)} & \textbf{Speedup} \\
    \hline\hline
    Actor1         & 0 & 11 & 77  & 20 & 3.9$\times$ \\
    Actor1         & 3 & 56 & 100 & 28 & 3.6$\times$ \\
    flame\_salmon  & 0 & 11 & 101 & 31 & 3.3$\times$ \\
    flame\_salmon  & 2 & 35 & 125 & 57 & 2.2$\times$ \\
    sear\_steak    & 0 & 11 & 79  & 25 & 3.2$\times$ \\
    sear\_steak    & 2 & 35 & 96  & 47 & 2.0$\times$ \\
    \hline
    \end{tabular}
  }
  \caption{RAHT decode latency on Jetson Orin (ms). CUDA vs.\ PyTorch implementation. Averaged across sampled frames.}
  \label{tab:jetson-raht-decode}
\end{table}

\begin{table}[b]
  \centering
  \footnotesize
  \begin{tabular}{l|rrr|rr}
    \hline
    & \multicolumn{3}{c|}{\textbf{Compressed Size (MB)}} & \multicolumn{2}{c}{\textbf{Relative Ratio}} \\
    \textbf{Sequence} & RGB & YUV & KLT & RGB/KLT & YUV/KLT \\
    \hline\hline
    flame\_salmon  & 11.94 & 11.37  & 8.35  & 1.43  & 1.36  \\
    sear\_steak    & 9.29 & 8.99  & 6.78    & 1.37  & 1.33  \\
    Actor1         & 10.97 & 10.43  & 9.90  & 1.11  & 1.05  \\
    \hline
  \end{tabular}
  \caption{Impact of color decorrelation on compressed size (megabytes).}
  \label{tab:ablation-klt}
\end{table}

\end{document}